\documentclass[12pt]{iopart}
\usepackage{graphicx}
\usepackage{booktabs}
\usepackage{bm}
\usepackage{subfigure}

\begin{document}

\title[Testing gravitational redshift using TFC method for ACES frequency links]{Formulation to test gravitational redshift based on the tri-frequency combination of ACES frequency links}
\author{X. Sun$^1$ $^2$, W. B. Shen$^1$ $^2$\footnote{Corresponding authors: Wen-Bin Shen (wbshen@sgg.whu.edu.cn)}, Z. Shen$^3$, C. Cai$^1$, W. Xu$^1$ and P. Zhang$^1$}
\address{$^1$ Department of Geophysics, School of Geodesy and Geomatics/Key Laboratory of Geospace Environment and Geodesy of Ministry of Education, Wuhan University, Wuhan, China}
\address{$^2$ State Key Laboratory of Information Engineering in Surveying, Mapping and Remote Sensing, Wuhan University, Wuhan, China}
\address{$^3$ Hubei University of Science and Technology, Xianning, China}

\begin{abstract}
Atomic Clock Ensemble in Space (ACES) is an ESA mission mainly designed to test gravitational redshift with high-performance atomic clocks in space and on the ground. A crucial part of this experiment lies in its two-way Microwave Link (MWL), which uses the uplink of carrier frequency 13.475 GHz (Ku band) and downlinks of carrier frequencies 14.70333 GHz (Ku band) and 2248 MHz (S band) to transfer time and frequency. The formulation based on the time comparison has been studied for over a decade. However, there are advantages of using frequency comparison instead of time comparison to test gravitational redshift. Hence, we develop a tri-frequency combination (TFC) method based on the measurements of the frequency shifts of three independent MWLs between ACES and a ground station. The potential scientific object requires stabilities of atomic clocks   at least $3\times10^{-16}$/day, so we must consider various effects, including the Doppler effect, second-order Doppler effect, atmospheric frequency shift, tidal effects, refraction caused by the atmosphere, and Shapiro effect, with accuracy levels of tens of centimeters. The ACES payload will be launched as previously planned in the middle of 2021, and the formulation proposed in this study will enable testing gravitational redshift at an accuracy level of at least $2\times10^{-6}$, which is more than one order higher than the present accuracy level of $7\times10^{-5}$.
\end{abstract}
\noindent{\it Keywords\/}:ACES mission, gravitational redshift, tri-frequency combination, frequency comparison, microwave links

\submitto{\MET}
\maketitle

\section{Introduction}
\label{sec:1}
General relativity theory \cite{Einstein1915} concludes three classic predictions: Mercury precession, light deflection and gravitational redshift. The first two have been confirmed by \cite{Einstein1915} and a group led by \cite{Eddington1919}, but the gravitational redshift was not tested until 1960.

The first direct experimental verifications of gravitational redshift are the series of Pound-Rebka-Snider experiments during 1960--1965 \cite{Pound1960,Pound1965}, who observed the shift using a Mössbauer emitter and absorber at the Jefferson Physical Laboratory tower at Harvard University. Later, there is an around-the-world experiment. Four cesium beam clocks were used to fly around the world on commercial jet flights during several days in October 1971, and they flew in opposite directions while recording the time differences \cite{Hafele1972}. Additionally, other types of experiments measure the shift of spectral lines in the Sun's gravitational field since 1960 \cite{Brault1962}. Typically, a Galileo solar redshift experiment tested the gravitational redshift to 1\% accuracy \cite{Krisher1993}. The most famous test was obtained by the Gravity Probe A (GPA) mission in June 1976, which launched a hydrogen maser onboard a rocket to a height of 10\,000 km \cite{Vessot1979}. During its flight, frequency comparisons were conducted between the maser on the rocket and a corresponding maser on the ground. The consistency of the relativistic frequency shift with the prediction was $7\times10^{-5}$ \cite{Vessot1980}. Until now, the most precise indirect tests were performed by eccentric Galileo satellites. The tests are based on the satellites GSAT-0201 and GSAT-0202 of the European Global Navigation Satellite System (GNSS) Galileo, which were accidentally delivered on elliptic instead of circular orbits. Two research teams simultaneously published their results with $(0.19\pm2.48)\times10^{-5}$ \cite{Delva2018} and $(-0.9\pm1.4)\times10^{-5}$ \cite{Herrmann2018}, respectively.

The Atomic Clock Ensemble in Space (ACES) experiment \cite{Cacciapuoti2009,Cacciapuoti2011,Cacciapuoti2017,Meynadier2018}, which was installed onboard the International Space Station (ISS), is an ESA-CNES mission mainly planned to test the gravitational redshift. Equipped with atomic clocks of fractional frequency instability and inaccuracy of $(1-3)\times10^{-16}$, it aims to test the gravitational redshift at a level of $2\times10^{-6}$ \cite{Cacciapuoti2011,Cacciapuoti2017}, which is one and a half orders higher than the GPA experiment.

The main onboard instruments are an active hydrogen maser (SHM) and a cold cesium atoms (PHARAO). The PHARAO clock reaches a fractional frequency stability of $1.1\times10^{-13}\sqrt{\tau}$, where $\tau$ is the integration time in seconds, and an accuracy of a few parts in $10^{16}$ \cite{Cacciapuoti2017}. Meanwhile, SHM demonstrates a fractional frequency instability of $1.5\times10^{-15}$ after 10\,000 s of integration time. Combining the short-term stability of the H-maser with the long-term stability and accuracy of the cesium clock, two clocks will generate an on-board time scale \cite{Cacciapuoti2011,Cacciapuoti2017}. 

ACES enables frequency/time comparisons between ISS and ground stations by using two independent time \& frequency transfer links (Microwave Links (MWL) and European Laser Timing (ELT) optical link) to test general relativity and develop applications in geodesy (relativistic geodesy) and time \& frequency metrology \cite{Cacciapuoti2011,Cacciapuoti2017}. These science objectives are closely related to the MWL performance \cite{Meynadier2018}, and its performance plays a key role in this study. MWL uses the uplink of carrier frequency 13.475 GHz (Ku band) and downlinks of carrier frequencies 14.70333 GHz (Ku band) and 2248 MHz (S band) to transfer time and frequency. MWL will perform with time deviation better than 0.3 ps at 300 s, 7 ps at 1 day, and 23 ps at 10 days of integration time \cite{Hess2011}. These performances, which surpass those of existing techniques (TWSTFT and GPS) by 1--2 orders of magnitude, will enable comparisons (common view and uncommon view) of ground clocks with $10^{-17}$ frequency resolution after a few days of integration \cite{Hess2011}.

Concerning the ACES mission, some studies have addressed the test of gravitational redshift based on time comparison \cite{Cacciapuoti2009,Duchayne2009,Meynadier2018}, but there are almost no publications related to the frequency comparison. Compared with time comparison, frequency comparison has the following advantages: (1) it can weaken the effect of the phase ambiguity because the frequency measurement is irrelevant with ranging and is a consequence of counting during a short time; (2) it can determine the instant gravitational potential, while for time comparison, we must accumulate data to solve the time changing rate to deduce the gravitational redshift value. However, the accuracy of measuring the instant frequency is largely constrained, which implies that we must also accumulate observations to obtain results with higher accuracy.

In our study, we proposed a new formulation, referred to as tri-frequency combination (TFC) to obtain the gravitational potential difference by combining three frequency observations. For the one-way frequency transfer model with a precision requirement of $10^{-16}$, we adopt a formulation accurate to $c^{-3}$ order in free space with medium, which was proposed by \cite{Blanchet2001}. For our theoretical contributions, we extended the model of \cite{Blanchet2001} from free space (vacuum) to real space with media (see section \ref{sec:2} and \ref{sec:a}) and formulated the approach to eliminate the Doppler frequency shift (the term Doppler effect or Doppler frequency shift mentioned in this paper refers to the first-order Doppler effect) considering the time offset among three links (see section \ref{sec:3} and \ref{sec:b}). Our final TFC model can successfully eliminate all types of shifts to the order of $10^{-16}$. To verify our model and analyze the demanded magnitude of parameters, we designed simulation experiments considering the real orbit, reliable clocks noises, real atmosphere and real gravity (see section \ref{sec:5}).

\section{One-way frequency transfer between ISS and ground station}
\label{sec:2}
For MWLs, the ACES mission uses two different antennas: one Ku-band antenna for uplink and downlink and one S-band antenna for only downward signals. It uses the uplink of carrier frequency 13.475 GHz (Ku band, and the frequency shift will be broadcast to the ground station afterwards) and downlinks of carrier frequency 14.70333 GHz (Ku band) and 2248 MHz (S band) \cite{Cacciapuoti2009}. These three frequencies are denoted by $f_\mathrm{1}$, $f_\mathrm{2}$, and $f_\mathrm{3}$ throughout this study. The goal of testing accuracy is $2\times10^{-6}$; thus, we need a frequency transfer model to the level of $1\times10^{-16}$, which requires a relativistic model to the order of $c^{-3}$. 

First, we consider a downlink from satellite A to ground B. The frequency transfer ratio $f_\mathrm{A}/f_\mathrm{B}$ between proper frequencies $f_\mathrm{A}$ and $f_\mathrm{B}$ is determined by the clocks on the satellite (A) and the ground (B). In practice, this is achieved using the transmission of photons from A to B and the following formula
\begin{equation}\label{fab}
\frac{{{f}_\mathrm{A}}}{{{f}_\mathrm{B}}}=\left(\frac{{{f}_\mathrm{A}}}{{{\nu}_\mathrm{A}}}\right)\left(\frac{{{\nu}_\mathrm{A}}}{{{\nu}_\mathrm{B}}}\right)\left(\frac{{{\nu}_\mathrm{B}}}{{{f}_\mathrm{B}}}\right)
\end{equation}
where $\nu_\mathrm{A}$ and $\nu_\mathrm{B}$ are the proper frequencies of the photon at A and B. In a general relativistic framework, the proper frequency shift of the photon from A to B is expressed by \cite{Blanchet2001}
\begin{equation}\label{nuba}
\frac{{{\nu _\mathrm B}}}{{{\nu _\mathrm A}}} = \frac{{1 - \frac{1}{{{c^2}}}\left[ {{U_\mathrm E}\left( {{{\vec{r}}_\mathrm A}} \right) + \frac{{v_\mathrm A^2}}{2}} \right]}}{{1 - \frac{1}{{{c^2}}}\left[ {{U_\mathrm E}\left( {{{\vec{r}}_\mathrm B}} \right) + \frac{{v_\mathrm B^2}}{2}} \right]}}\frac{{{q_\mathrm B}}}{{{q_\mathrm A}}}
\end{equation}
The first factor on the right-hand side is the sum of the gravitational redshift and transverse Doppler frequency shift, and $U_\mathrm{E}$ is Newtonian potential of the Earth in the frame of Earth-Centered Earth-Fixed (ECEF). We denote radial vectors ${\vec r}_\mathrm A={\vec x}_\mathrm A\left(t_\mathrm A\right)$ and ${\vec r}_\mathrm B={\vec x}_\mathrm B\left(t_\mathrm B\right)$, so $r_\mathrm A=\left|{\vec r}_\mathrm A\right|$, $r_\mathrm B=\left|{\vec r}_\mathrm B\right|$. ${\vec v}_\mathrm A={\vec v}_\mathrm A\left(t_\mathrm A\right)$ and ${\vec v}_\mathrm B={\vec v}_\mathrm B\left(t_\mathrm B\right)$ are the coordinate velocities. To the required order of $1/c^3$, the last factor in equation (\ref{nuba}) is obtained from \cite{Blanchet2001}
\begin{equation}\label{qa}
{q_\mathrm A} = 1 - \frac{{{{\vec{N}}_\mathrm{AB}} \cdot {{\vec{v}}_\mathrm A}}}{c} - \frac{{4G{M_\mathrm E}}}{{{c^3}}}\frac{{\left( {{r_\mathrm A} + {r_\mathrm B}} \right){{\vec{N}}_\mathrm{AB}} \cdot {{\vec{v}}_\mathrm A} + {R_\mathrm{AB}}\frac{{{{\vec{r}}_\mathrm A} \cdot {{\vec{v}}_\mathrm A}}}{{{r_\mathrm A}}}}}{{{{\left( {{r_\mathrm A} + {r_\mathrm B}} \right)}^2} - R_\mathrm{AB}^2}}
\end{equation}
\begin{equation}\label{qb}
{q_\mathrm B} = 1 - \frac{{{{\vec{N}}_\mathrm{AB}} \cdot {{\vec{v}}_\mathrm B}}}{c} - \frac{{4G{M_\mathrm E}}}{{{c^3}}}\frac{{\left( {{r_\mathrm A} + {r_\mathrm B}} \right){{\vec{N}}_\mathrm{AB}} \cdot {{\vec{v}}_\mathrm B} - {R_\mathrm{AB}}\frac{{{{\vec{r}}_\mathrm B} \cdot {{\vec{v}}_\mathrm B}}}{{{r_\mathrm B}}}}}{{{{\left( {{r_\mathrm A} + {r_\mathrm B}} \right)}^2} - R_\mathrm{AB}^2}}
\end{equation}
with ${\vec R}_\mathrm{AB}={\vec r}_\mathrm B-{\vec r}_\mathrm A$, $R_\mathrm{AB}=|{\vec R}_\mathrm{AB}|$, and ${\vec N}_\mathrm{AB}={\vec R}_\mathrm{AB}/R_\mathrm{AB}$. The last terms in equations (\ref{qa}) and (\ref{qb}) are caused by curved geometry in the general relativistic framework and referred to as Shapiro effect. In this formulation, we approximate the Earth as a spherically symmetric body, since the J2 term does not exceed the magnitude of $4\times10^{-17}$ \cite{Blanchet2001}.

Using the simplified notation
\begin{equation}\label{simplify}
\cases{
{A_\mathrm{Shap}} = \frac{{4G{M_\mathrm E}}}{{{c^3}}}\left[ \frac{{\left( {{r_\mathrm A} + {r_\mathrm B}} \right){{\vec{N}}_\mathrm{AB}} \cdot {{\vec{v}}_\mathrm A} + {R_\mathrm{AB}}\frac{{{{\vec{r}}_\mathrm A} \cdot {{\vec{v}}_\mathrm A}}}{{{r_\mathrm A}}}}}{{{{\left( {{r_\mathrm A} + {r_\mathrm B}} \right)}^2} - R_\mathrm{AB}^2}} \right.\\
\qquad \qquad \left.- \frac{{\left( {{r_\mathrm A} + {r_\mathrm B}} \right){{\vec{N}}_\mathrm{AB}} \cdot {{\vec{v}}_\mathrm B} - {R_\mathrm{AB}}\frac{{{{\vec{r}}_\mathrm B} \cdot {{\vec{v}}_\mathrm B}}}{{{r_\mathrm B}}}}}{{{{\left( {{r_\mathrm A} + {r_\mathrm B}} \right)}^2} - R_\mathrm{AB}^2}} \right]\\
{A_\mathrm{rel}} = \frac{{1 - \frac{1}{{{c^2}}}\left[ {{U_\mathrm E}\left( {{r_\mathrm A}} \right) + \frac{{v_\mathrm A^2}}{2}} \right]}}{{1 - \frac{1}{{{c^2}}}\left[ {{U_\mathrm E}\left( {{r_\mathrm B}} \right) + \frac{{v_\mathrm B^2}}{2}} \right]}}\\
{A_\mathrm{dop}} = \frac{{1 - \frac{{{{\vec{N}}_\mathrm{AB}} \cdot {{\vec{v}}_\mathrm B}}}{c}}}{{1 - \frac{{{{\vec{N}}_\mathrm{AB}} \cdot {{\vec{v}}_\mathrm A}}}{c}}}
}
\end{equation}
we have
\begin{equation}\label{nuba_simplify}
\frac{{{\nu _\mathrm B}}}{{{\nu _\mathrm A}}} = {A_\mathrm{rel}}\left( {{A_\mathrm{dop}} + {A_\mathrm{Shap}}} \right)
\end{equation}
where $A_\mathrm{Shap}$ is Shapiro effects given by equations (\ref{qa}) and (\ref{qb}).

\begin{figure}  
\centering  
\includegraphics[width=10cm]{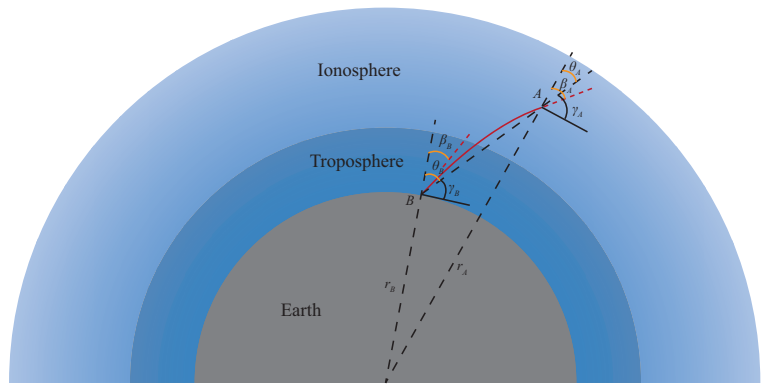}  
\caption{Principle of ray refraction through the atmosphere for ISS. $r$ is distance from the Earth center, $\beta$ is the angle between the tangent of the electromagnetic wave and the normal of the layer, $\theta$ is the angle between AB line direction and layer normal, and $\gamma$ is the complementary angle of $\theta$.}
\label{fig1}  
\end{figure}

Models (\ref{simplify})-(\ref{nuba_simplify}) only hold in vacuum. In a real space with medium, electromagnetic waves experience a change in direction of propagation or refractive bending when they are transmitted through the atmosphere, which is divided into the troposphere (0--60 km) and ionosphere (60--2000 km). Because of the refraction phenomenon, the direction of the refracted ray at the space station slightly differs from the unrefracted line-of-sight direction \cite{Millman1984}, as figure \ref{fig1} shows. This phenomenon has significant applications in the GPS/MET (Global Positioning System/Meteorology) experiment \cite{Hajj1998} and will cause a slight change of Doppler effect in equation (\ref{nuba_simplify}). Here, by defining ${\bar A}_\mathrm{dop}$ as Doppler frequency considering the atmosphere, we have
\begin{equation}\label{nuba_correct}
\frac{{{\nu _\mathrm B}}}{{{\nu _\mathrm A}}} = {A_\mathrm{rel}}\left( {{{\bar A}_\mathrm{dop}} + {A_\mathrm{Shap}}} \right)
\end{equation}

The refractive index in the ionosphere is relevant with the carrier frequency, but the refractive index in troposphere is nearly irrelevant with it. Therefore, for all links $f_1$, $f_2$ and $f_3$, supposing that they are simultaneously emitted, the bending effects of the troposphere are approximately identical, which makes it easy to wipe out the tropospheric part. For the ionospheric part, to the order of $f^{-2}$, we have \cite{Davies1962,Hajj1998}

\begin{equation}\label{n}
n = 1 - 40.3\frac{{{n_\mathrm e}}}{{{f^2}}}
\end{equation}
where $n_\mathrm E$ is the electron density per cubic meter; high orders such as $f^{-3}$ are neglected because they are at least two magnitudes smaller than the order of $f^{-2}$ \cite{Petit2010}, which we will later analyze.

In the phase form, Doppler frequency shift is given by phase path $P$ of the radio wave \cite{Davies1962,Davies1966,Jacobs1966,Shen2016}:
\begin{equation}\label{dfdop}
\Delta {f_\mathrm{dop}} =  - \frac{f}{c}\frac{{dP}}{{dt}} =  - \frac{1}{\lambda }\frac{{dP}}{{dt}}
\end{equation}
where the velocity of the source induces a change in $\lambda$, and the velocity of the observer changes $dP/dt$. Thus, in vacuum, the first-order Doppler frequency shift can be derived in terms of $A_\mathrm{dop}$, as expressed by equation (\ref{simplify}).

More specifically, we have \cite{Bennett1968}
\begin{equation}\label{dfdop}
\Delta {f_\mathrm{dop}} =  - \frac{1}{\lambda }\left( {\int_\mathrm A^B {\frac{{\partial n}}{{\partial t}}\cos \alpha ds}  + {n_\mathrm B}{{\vec{T}}_\mathrm B} \cdot {{\vec{v}}_\mathrm B} - {n_\mathrm A}{{\vec{T}}_\mathrm A} \cdot {{\vec{v}}_\mathrm A}} \right)
\end{equation}
where ${\vec T}$ is a unit vector of the wave normal, $n$ is the refractive index, and $\alpha$ is the angle between the wave normal and the ray direction. We suppose that the atmosphere is an isotropic medium, the refractive index is nearly independent of ray directions, and $\alpha=0$ \cite{Davies1965}. From this equation, the atmospheric influence can be explained by two reasons: the refractive index varies with time \cite{Davies1966,Jacobs1966}, and the wave path varies because the observer moves \cite{Melbourne1994}.

Due to the velocity of the source, wavelength $\lambda$ is expressed as
\begin{equation}\label{lambda}
\lambda  = {\lambda _0}\left( {1 - \frac{{{n_\mathrm A}{{\vec{T}}_\mathrm A} \cdot {{\vec{v}}_\mathrm A}}}{c}} \right)
\end{equation}
Considering ${\bar A}_\mathrm{dop}$ defined in equation (\ref{nuba_correct}), with (\ref{dfdop}) and (\ref{lambda}), we have
\begin{equation}\label{adopbar}
{\bar A_\mathrm{dop}} = \frac{{f + \Delta {f_\mathrm{dop}}}}{f} = \frac{{1 - \frac{{{n_\mathrm B}{{\vec{T}}_\mathrm B} \cdot {{\vec{v}}_\mathrm B}}}{c} - \int_\mathrm A^B {\frac{{\partial n}}{{\partial t}}\cos \alpha ds} }}{{1 - \frac{{{n_\mathrm A}{{\vec{T}}_\mathrm A} \cdot {{\vec{v}}_\mathrm A}}}{c}}}
\end{equation}
With the height of ISS of approximately 400 km \cite{Cacciapuoti2009}, the ACES-ground links lie in the middle layer of the ionosphere. The integral term of the refractive index in equation (\ref{adopbar}) can be expressed as the sum of the ionospheric and tropospheric parts, and if we suppose $\alpha=0$, we have \cite{Shen2016}
\begin{equation}\label{inteindex}
\int_\mathrm A^B {\frac{{\partial n}}{{\partial t}}\cos \alpha ds}  =  - \frac{{40.3}}{{c{f^2}}}\frac{d}{{dt}}\int_\mathrm{Li} {{n_\mathrm e}ds}  + \frac{1}{c}\frac{d}{{dt}}\int_\mathrm{Lt} {\left( {{M_1} + {M_2}} \right)ds}
\end{equation}
where ${n_\mathrm e}$ is the electron density along the trajectory, ${M_1} = 77.6 \times {10^{-6}}p/T$, and ${M_2} = 0.373\varepsilon /{T^2}$ with temperature $T$, total pressure $p$ and partial pressure of water vapor $\varepsilon$ along the trajectory. 

For this expansion, equation (\ref{inteindex}) can be rewritten as
\begin{equation}\label{adopbar2}
{\bar A_\mathrm{dop}} = \frac{{1 - \frac{{{n_\mathrm B}{{\vec{T}}_\mathrm B} \cdot {{\vec{v}}_\mathrm B}}}{c}}}{{1 - \frac{{{n_\mathrm A}{{\vec{T}}_\mathrm A} \cdot {{\vec{v}}_\mathrm A}}}{c}}}+\frac{{40.3}}{{c{f^2}}}\frac{d}{{dt}}\int_\mathrm{Li} {{n_\mathrm e}ds}  - \frac{1}{c}\frac{d}{{dt}}\int_\mathrm{Lt} {\left( {{M_1} + {M_2}} \right)ds}
\end{equation}
where (referring to \ref{sec:a})
\begin{equation}\label{appendixa}
\frac{{1 - \frac{{{n_\mathrm B}{{\vec{T}}_\mathrm B} \cdot {{\vec{v}}_\mathrm B}}}{c}}}{{1 - \frac{{{n_\mathrm A}{{\vec{T}}_\mathrm A} \cdot {{\vec{v}}_\mathrm A}}}{c}}} = \frac{{1 - \frac{{{{\vec{N}}_\mathrm{AB}} \cdot {{\vec{v}}_\mathrm B}}}{c} + \frac{{{v_\mathrm{Bx}}{\delta _\mathrm B}\sin {\gamma _\mathrm B} - {v_\mathrm{By}}{\delta _\mathrm B}\cos {\gamma _\mathrm B}}}{c} - \frac{{\left( {{M_1} + {M_2}} \right){{\vec{N}}_\mathrm{AB}} \cdot {{\vec{v}}_\mathrm B}}}{c}}}{{1 - \frac{{{{\vec{N}}_\mathrm{AB}} \cdot {{\vec{v}}_\mathrm A}}}{c} - \frac{{{v_\mathrm{Ax}}{\delta _\mathrm A}\sin {\gamma _\mathrm B} - {v_\mathrm{Ay}}{\delta _\mathrm A}\cos {\gamma _\mathrm B}}}{c} + \frac{{40.3{n_\mathrm e}{{\vec{N}}_\mathrm{AB}} \cdot {{\vec{v}}_\mathrm A}}}{{c{f^2}}}}}
\end{equation}
where $v_\mathrm{Ax}$, $v_\mathrm{Ay}$, $v_\mathrm{Bx}$, and $v_\mathrm{By}$ are components of velocities $\vec{v}_\mathrm A$ and $\vec{v}_\mathrm B$ of space station A and ground site B projected in the refraction plane in figure \ref{fig1}, and $\delta_\mathrm A$ and $\delta_\mathrm B$ are deviated angles from the line of sight. More details are referred to \ref{sec:a}.

Focusing on the effect of the variation of the refractive index and wave path, we can obtain the expression of ${\bar A_\mathrm{dop}}$; then, we separate the refraction part (not time-varying parts), ionospheric and tropospheric frequency shift (time-varying parts), and Shapiro effect. Thus, the one-way frequency transfer model should be written as
\begin{equation}\label{nuba2}
\frac{{{\nu _\mathrm B}}}{{{\nu _\mathrm A}}} = {A_\mathrm{rel}}\left( {{A_\mathrm{dop}} + \delta {f_\mathrm{refr}} + \delta {f_\mathrm{ion}} + \delta {f_\mathrm{trop}} + {A_\mathrm{Shap}}} \right)
\end{equation}
where
\begin{equation}\label{simplify2}
\eqalign{
\delta {f_\mathrm{refr}} &= \frac{{\left( {{v_\mathrm{Ax}}{\delta _\mathrm A} + {v_\mathrm{Bx}}{\delta _\mathrm B}} \right)\sin {\gamma _\mathrm B} - \left( {{v_\mathrm{Ay}}{\delta _\mathrm A} + {v_\mathrm{By}}{\delta _\mathrm B}} \right)\cos {\gamma _\mathrm B}}}{c} \cr
&- \frac{{\left( {{M_1} + {M_2}} \right){{\vec{N}}_\mathrm{AB}} \cdot {{\vec{v}}_\mathrm B}}}{c} - \frac{{40.3{n_\mathrm e}{{\vec{N}}_\mathrm{AB}} \cdot {{\vec{v}}_\mathrm A}}}{{c{f^2}}}\cr
\delta {f_\mathrm{ion}} &= \frac{{40.3}}{{c{f^2}}}\frac{d}{{dt}}\int_\mathrm{Li} {\frac{{d{n_\mathrm e}}}{{dt}}ds} \cr
\delta {f_\mathrm{trop}} &= - \frac{1}{c}\frac{d}{{dt}}\int_\mathrm{Lt} {\frac{{d\left( {{M_1} + {M_2}} \right)}}{{dt}}ds} 
}
\end{equation}
where $\delta {f_\mathrm{refr}}$ is the bending effect on Doppler frequency shift, which is caused by refraction, $\delta {f_\mathrm{ion}}$ and $\delta {f_\mathrm{trop}}$ are atmospheric effects caused by the time-varying refractive index. For the ACES links, estimates show that the magnitude of $\delta f_\mathrm{dop}$ is approximately $10^{-5}$; $\delta f_\mathrm{rel}$ is approximately $10^{-10}$ \cite{Blanchet2001}; $\delta f_\mathrm{ion}$, $\delta f_\mathrm{ion}$ and $\delta f_\mathrm{trop}$ will be estimated in section \ref{sec:5}.

\section{Formulation to test gravitational redshift}
\label{sec:3}
\subsection{Tri-frequency combination}
\label{sec:3_1}
\begin{figure}  
\centering  
\includegraphics[width=8cm]{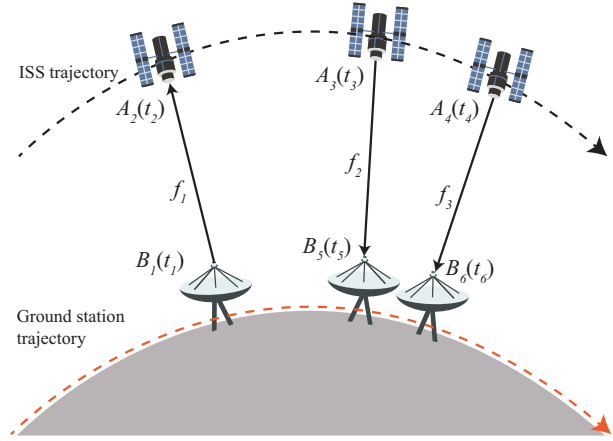}  
\caption{MWL principle (modified after \cite{Meynadier2017}). At time $t_1$, the ground station emits signal $f_1$ to ISS, which is received at time $t_2$. Meanwhile, two signals $f_2$ and $f_3$ are emitted to the ground station at time $t_3$ and $t_4$ (they are approximately $t_2$), which are received at time $t_5$ and $t_6$.}
\label{fig2}  
\end{figure}

Although all of these links are independent and can be synchronized afterwards by data processing, the synchronization error may cause severe residual errors. In this study, we list coordinate time $t_1\sim t_6$ to identify six events and use a combination of three frequency links of ACES to test the gravitational redshift, as figure \ref{fig2} shows. Later, we will analyze the required synchronization precision to test gravitational redshift at the $2\times10^{-6}$ level.

In our formulation, we use a nonrotating geocentric space-time coordinate system. At coordinate time $t_1$, Ku-band signal $f_1$ is emitted and received by the space station at coordinate time $t_2$. Meanwhile, two signals $f_2$ and $f_3$ are emitted from the space station at coordinate time $t_3$ and $t_4$, respectively, and received by the ground station at coordinate time $t_5$ and $t_6$, respectively. If we define a coordinate time interval by $T_{ij}=t_j-t_i$, $T_{23}$ and $T_{34}$ will theoretically be synchronized to zero, but in practice, there is a difference between them.

For ACES links $f_1=13.475$ GHz, $f_2=14.70333$ GHz and $f_3=2248$ MHz, the third link is of low frequency, which greatly suffers from ionospheric effects. If we suppose that $T_{34}$ is extremely small ($<1\ \mathrm{\mu s}$), the only different error between link 2 and link 3 is the ionospheric error, because other shifts in link 2 and link 3 are close. We define $f_1^{\prime}$, $f_2^{\prime}$ and $f_3^{\prime}$ as the received frequencies corresponding to emitted frequencies $f_1$, $f_2$ and $f_3$.

If we divide frequency shift $f_2^{\prime}/f_2$ by frequency shift $f_3^{\prime}/f_3$, based on equations (\ref{simplify}), (\ref{nuba2}) and (\ref{simplify2}), the Doppler part, relativistic parts (hereafter, they refer to transverse Doppler effects and gravitational redshift) and tropospheric part are cancelled, and we obtain 
\begin{eqnarray}\label{f23}
\fl {{\frac{{{f_2}'}}{{{f_2}}}} \mathord{\left/
 {\vphantom {{\frac{{{f_2}'}}{{{f_2}}}} {\frac{{{f_3}'}}{{{f_3}}}}}} \right.
 \kern-\nulldelimiterspace} {\frac{{{f_3}'}}{{{f_3}}}}} = 1 + \left( {1 - \frac{{f_2^2}}{{f_3^2}}} \right)\left[ {\frac{{40.3}}{{cf_2^2}}\frac{d}{{dt}}\int_\mathrm{Li} {\frac{{d{n_\mathrm e}}}{{dt}}ds}  + \frac{{\left( {{v_\mathrm{Ax}}\delta _\mathrm A^\mathrm{ion} + {v_\mathrm{Bx}}\delta _\mathrm B^\mathrm{ion}} \right)\sin {\gamma _\mathrm B}}}{c}} \right.\nonumber\\
\qquad \left. { - \frac{{\left( {{v_\mathrm{Ay}}\delta _\mathrm A^\mathrm{ion} + {v_\mathrm{By}}\delta _\mathrm B^\mathrm{ion}} \right)\cos {\gamma _\mathrm A}}}{c} + \frac{{40.3{n_\mathrm e}{{\vec{N}}_\mathrm{AB}} \cdot {{\vec{v}}_\mathrm A}}}{{cf_2^2}}} \right]
\end{eqnarray}
where $\delta _\mathrm A^\mathrm{ion}$ and $\delta _\mathrm B^\mathrm{ion}$ are the refractive angles relevant with the carrier frequency, and they are inversely proportional to the square of the carrier frequency (see \ref{sec:a}). All terms in the last middle bracket in equation (\ref{f23}) are inversely proportional to the square of the carrier frequency.

Referring to figure \ref{fig2}, the ground station has moved a certain distance (approximately 1 meter) from $t_1$ to $t_5$; thus, the first-order Doppler frequency shift cannot be completely cancelled from equation (\ref{nuba2}). By the same technique, we divide frequency shift $f_1^{\prime}/f_1$ by frequency shift $f_2^{\prime}/f_2$ to obtain
\begin{equation}\label{f12}
{{\frac{{{f_1}'}}{{{f_1}}}} \mathord{\left/
 {\vphantom {{\frac{{{f_1}'}}{{{f_1}}}} {\frac{{{f_2}'}}{{{f_2}}}}}} \right.
 \kern-\nulldelimiterspace} {\frac{{{f_2}'}}{{{f_2}}}}} = \frac{{{m_1}}}{{{m_2}}}\frac{{\left[ {\frac{{1 - \frac{1}{{{c^2}}}\left( {{U_\mathrm{B1}} + \frac{{v_\mathrm{B1}^2}}{2}} \right)}}{{1 - \frac{1}{{{c^2}}}\left( {{U_\mathrm{A2}} + \frac{{v_\mathrm{A2}^2}}{2}} \right)}}\left( {\frac{{1 - \frac{{{{\vec{N}}_\mathrm{B1A2}} \cdot {{\vec{v}}_\mathrm{A2}}}}{c}}}{{1 - \frac{{{{\vec{N}}_\mathrm{B1A2}} \cdot {{\vec{v}}_\mathrm{B1}}}}{c}}} - {A_\mathrm{Shap,link1}}} \right)} \right]}}{{\left[ {\frac{{1 - \frac{1}{{{c^2}}}\left( {{U_\mathrm{A3}} + \frac{{v_\mathrm{A3}^2}}{2}} \right)}}{{1 - \frac{1}{{{c^2}}}\left( {{U_\mathrm{B5}} + \frac{{v_\mathrm{B5}^2}}{2}} \right)}}\left( {\frac{{1 - \frac{{{{\vec{N}}_\mathrm{A3B5}} \cdot {{\vec{v}}_\mathrm{B5}}}}{c}}}{{1 - \frac{{{{\vec{N}}_\mathrm{A3B5}} \cdot {{\vec{v}}_\mathrm{A3}}}}{c}}} - {A_\mathrm{Shap,link2}}} \right)} \right]}}
\end{equation}
where $Bi$ is the position of the ground station at time $t_i$, and $Ai$ is the position of ISS at time $t_i$. Here, $m_1/m_2$ is the solved ionospheric part
\begin{eqnarray}\label{m12}
\fl \frac{{{m_1}}}{{{m_2}}} = 1 + \left( {\frac{{f_2^2}}{{f_1^2}} - 1} \right)\left[ {\frac{{40.3}}{{cf_2^2}}\frac{d}{{dt}}\int_\mathrm{Li} {\frac{{d{n_\mathrm e}}}{{dt}}ds}  + \frac{{\left( {{v_\mathrm{Ax}}\delta _\mathrm A^\mathrm{ion} + {v_\mathrm{Bx}}\delta _\mathrm B^\mathrm{ion}} \right)\sin {\gamma _\mathrm B}}}{c}} \right.\nonumber\\
 \qquad \left. { - \frac{{\left( {{v_\mathrm{Ay}}\delta _\mathrm A^\mathrm{ion} + {v_\mathrm{By}}\delta _\mathrm B^\mathrm{ion}} \right)\cos {\gamma _\mathrm A}}}{c} + \frac{{40.3{n_\mathrm e}{{\vec{N}}_\mathrm{AB}} \cdot {{\vec{v}}_\mathrm A}}}{{cf_2^2}}} \right]{\kern 1pt} 
\end{eqnarray}
where the terms in the second bracket are related to ionospheric effects and determined by equation (\ref{f23}), which can give rise to the following relation
\begin{equation}\label{m12f}
\frac{{{m_1}}}{{{m_2}}} = 1 + \left( {\frac{{f_2^2}}{{f_1^2}} - 1} \right)\left( {\frac{{f_3^2}}{{f_3^2 - f_2^2}}} \right)\left( {{{\frac{{{f_2}'}}{{{f_2}}}} \mathord{\left/
 {\vphantom {{\frac{{{f_2}'}}{{{f_2}}}} {\frac{{{f_3}'}}{{{f_3}}}}}} \right.
 \kern-\nulldelimiterspace} {\frac{{{f_3}'}}{{{f_3}}}}} - 1} \right)
\end{equation}

If the difference in space parameters of the space station and ground station at different time points is ignored, the Doppler effect in equation (\ref{f12}) can be directly eliminated. However, at different time points such as $t_2$ and $t_3$, spatial parameters of ISS will be moderately different, which results in Doppler residuals in equation (\ref{f12}). According to \ref{sec:b} and accurate to the order of $c^{-3}$, we have
\begin{eqnarray}\label{tfc}
\fl {{\frac{{{f_1}'}}{{{f_1}}}} \mathord{\left/
 {\vphantom {{\frac{{{f_1}'}}{{{f_1}}}} {\frac{{{f_2}'}}{{{f_2}}}}}} \right.
 \kern-\nulldelimiterspace} {\frac{{{f_2}'}}{{{f_2}}}}}=\frac{{{m_1}}}{{{m_2}}}\left[ \frac{{1 - {{\left( {\frac{{{{\vec{N}}_\mathrm{A3B5}} \cdot {{\vec{v}}_\mathrm{A3}}}}{c}} \right)}^2} + \frac{{2\left( {{{\vec{N}}_\mathrm{A3B5}} \cdot {{\vec{v}}_\mathrm{A3}}} \right)\left( {{{\vec{N}}_\mathrm{A3B5}} \cdot {{\vec{v}}_\mathrm{B5}}} \right)}}{{{c^2}}} - \frac{{2{{\vec{v}}_\mathrm{A3}} \cdot {{\vec{v}}_\mathrm{B5}}}}{{{c^2}}} + {K_1}{T_{23}}}}{{1 + {{\left( {\frac{{{{\vec{N}}_\mathrm{A3B5}} \cdot {{\vec{v}}_\mathrm{B5}}}}{{{c^2}}}} \right)}^2} - \frac{{2v_\mathrm{B5}^2}}{{{c^2}}} - \frac{{2{{\vec{R}}_\mathrm{A3B5}} \cdot {{\vec{a}}_\mathrm{B5}}}}{{{c^2}}}+{K_2}{T_{23}}}} \right.\nonumber\\
\left.- {A_\mathrm{Shap,link2}} \right] \cdot {\left[ {1 - \frac{1}{{{c^2}}}\left( {{U_\mathrm{B5}} - {U_\mathrm{A3}} + \frac{{v_\mathrm{B5}^2 - v_\mathrm{A3}^2}}{2}} \right)} \right]^2}
\end{eqnarray}
\begin{equation}\label{tfc2}
\fl \eqalign{
{K_1} = \frac{{{{\vec{N}}_\mathrm{A3B5}} \cdot \left( {{{\vec{v}}_\mathrm{B5}} - {{\vec{v}}_\mathrm{A3}}} \right)\left( {{{\vec{N}}_\mathrm{A3B5}} \cdot {{\vec{v}}_\mathrm{A3}}} \right)}}{{c{R_\mathrm{A3B5}}}} - \frac{{\left( {{{\vec{v}}_\mathrm{B5}} - {{\vec{v}}_\mathrm{A3}}} \right) \cdot {{\vec{v}}_\mathrm{A3}}}}{{c{R_\mathrm{A3B5}}}} - \frac{{{{\vec{N}}_\mathrm{A3B5}} \cdot {{\vec{a}}_\mathrm{A3}}}}{c}\cr
{K_2} = \frac{{{{\vec{N}}_\mathrm{A3B5}} \cdot \left( {{{\vec{v}}_\mathrm{B5}} - {{\vec{v}}_\mathrm{A3}}} \right)\left( {{{\vec{N}}_\mathrm{A3B5}} \cdot {{\vec{v}}_\mathrm{B5}}} \right)}}{{c{R_\mathrm{A3B5}}}} - \frac{{\left( {{{\vec{v}}_\mathrm{B5}} - {{\vec{v}}_\mathrm{A3}}} \right) \cdot {{\vec{v}}_\mathrm{B5}}}}{{c{R_\mathrm{A3B5}}}}
}
\end{equation}
where $m_1/m_2$ is determined by equation (\ref{m12f}); ${A_\mathrm{Shap,link2}}$ is a term of $c^{-3}$; since the magnitudes of $T_{23}$ and $T_{15}$ are at most $c^{-1}$, the difference between ${A_\mathrm{Shap,link1}}$ and ${A_\mathrm{Shap,link2}}$ in equation (\ref{m12f}) is at most $c^{-4}$, which is negligible, as is the difference between $v_\mathrm{a2}^2/c^2$ and $v_\mathrm{a3}^2/c^2$.
\subsection{Error sources}
\label{sec:3_2}
Generally, errors are divided into systematic errors and random errors. All of the aforementioned errors are systematic errors, including Doppler frequency shift, atmospheric frequency shift (including ionospheric, tropospheric and refractive frequency shift), relativistic frequency shift and Shapiro frequency shift. These frequency shifts can be eliminated using our TFC model, but residuals remain. These residuals and tidal effects will be discussed in section \ref{sec:4_1}.

Random errors are caused by devices (e.g., atomic clocks, cables and emitters) and measurements (e.g., velocities and accelerations in equation (\ref{tfc}) and (\ref{tfc2})). Literatures \cite{Cacciapuoti2009,Cacciapuoti2017} have shown the Allan Deviation performance of ACES’s clocks (SHM and PHARAO), which shows that PHARAO has better long-term stability. However, these studies did not show the noise components of the clocks of ACES, and we can only simulate the clock data to approach their performance. For measurement noises, the accuracy for parameters ${\vec r}_\mathrm A$, ${\vec v}_\mathrm A$, ${\vec a}_\mathrm A$ and $T_{23}$ must be carefully controlled. Section \ref{sec:5} will discuss the parameter demands using our simulation data.

\section{Accuracy evaluation}
\label{sec:4}
\subsection{Residual errors}
\label{sec:4_1}
Although section \ref{sec:3_1} provided a practical calculation model to test gravitational redshift using the TFC method, there are residual errors, which are mainly reflected in two aspects: First, we have performed many approximations in the model derivation; Second, there are other types of errors in nature that we have not considered, such as tidal effects.

The model of the TFC method can be summarized by equations (\ref{m12f})-(\ref{tfc2}). In step one, frequency shifts of downlinks 2 and 3 are divided to obtain the ionospheric part; then, frequency shifts of uplink 1 and downlink 2 are divided. In step two, we substitute this result with the calculated residual Doppler effect and ionospheric part, so the gravitational potential difference can be calculated.
\subsubsection{Doppler residual errors}
\label{sec:4_1_1}
For Doppler residuals in section \ref{sec:3_1}, we only consider Doppler shift difference between link 1 and link 2 while ignoring that of link 2 and link 3. Since the time difference of link 2 and link 3 is less than $1\ \mathrm{\mu s}$ ($T_{34}<1 \ \mathrm{\mu s}$), and they are both downlinks, Doppler shift difference between link 2 and link 3 is much smaller. Supposing that $T_{34}$ is 100 ns, we take similar notes as section \ref{sec:3_1} and \ref{sec:b}: A denotes time $t_3$, B denotes time $t_5$, $B''$ denotes time $t_4$, and $A''$ denotes time $t_6$. Based on the spatial relation, we have
\begin{equation}\label{t56}
{T_{56}} = {T_{34}} - \frac{{{v_\mathrm A} \cdot {R_\mathrm{AB}}}}{{{c^2}}} + \frac{{{v_\mathrm B} \cdot {R_\mathrm{AB}}}}{{{c^2}}}
\end{equation}
With a numerical calculation with equation (\ref{t56}), we obtain that the numerical difference between $T_{34}$ and $T_{56}$ is tens of nanosecond, which implies that both $T_{34}$ and $T_{56}$ are in the order of magnitude of $c^{-2}$ and must be corrected. Through calculation, we neglected the intermediate process and obtained the relation of the Doppler shift difference between link 2 and link 3
\begin{eqnarray}\label{dop23}
\frac{{1 - \frac{{{{\vec{N}}_\mathrm{A''B''}} \cdot {{\vec{v}}_\mathrm{B''}}}}{c}}}{{1 - \frac{{{{\vec{N}}_\mathrm{A''B''}} \cdot {{\vec{v}}_\mathrm{A''}}}}{c}}} = \frac{{1 - \frac{{{{\vec{N}}_\mathrm{AB}} \cdot {{\vec{v}}_\mathrm B}}}{c}}}{{1 - \frac{{{{\vec{N}}_\mathrm{AB}} \cdot {{\vec{v}}_\mathrm A}}}{c}}}+\frac{{\left( {{{\vec{N}}_\mathrm{AB}} \cdot {{\vec{v}}_\mathrm B}} \right){{\left[ {{{\vec{N}}_\mathrm{AB}} \cdot \left( {{{\vec{v}}_\mathrm B} - {{\vec{v}}_\mathrm A}} \right)} \right]}^2}}}{{{c^3}}} \nonumber\\
- \frac{{\left( {{{\vec{v}}_\mathrm B} - {{\vec{v}}_\mathrm A}} \right) \cdot {{\vec{v}}_\mathrm B}{{\vec{N}}_\mathrm{AB}} \cdot \left( {{{\vec{v}}_\mathrm B} - {{\vec{v}}_\mathrm A}} \right)}}{{{c^3}}} - \frac{{\left( {{R_\mathrm{AB}} \cdot {a_\mathrm B}} \right){{\vec{N}}_\mathrm{AB}} \cdot \left( {{{\vec{v}}_\mathrm B} - {{\vec{v}}_\mathrm A}} \right)}}{{{c^3}}} + \nonumber\\
\left[ {\frac{{{{\left[ {{{\vec{N}}_\mathrm{AB}} \cdot \left( {{{\vec{v}}_\mathrm B} - {{\vec{v}}_\mathrm A}} \right)} \right]}^2}}}{{c{R_\mathrm{AB}}}} - \frac{{{{\left( {{{\vec{v}}_\mathrm B} - {{\vec{v}}_\mathrm A}} \right)}^2}}}{{c{R_\mathrm{AB}}}} - \frac{{{{\vec{N}}_\mathrm{AB}} \cdot \left( {{a_\mathrm B} - {a_\mathrm A}} \right)}}{c}} \right]{T_{34}}
\end{eqnarray}

With an approximate numerical calculation, $\left| {{\vec{N}_\mathrm{AB}} \cdot {\vec{v}_\mathrm A}/c} \right| \le 2.6 \times {10^{ - 5}}$, $\left| {{\vec{N}_\mathrm{AB}} \cdot {\vec{v}_\mathrm B}/c} \right| \le 1.6 \times {10^{ - 6}}$, $\left| {{{\vec{R}}_\mathrm{AB}} \cdot {{\vec{a}}_\mathrm A}/{c^2}} \right| \le 1.7 \times {10^{ - 10}}$, $\left| {{{\vec{R}}_\mathrm{AB}} \cdot {{\vec{a}}_\mathrm B}/{c^2}} \right| \le 7 \times {10^{ - 13}}$, and the largest part in equation (\ref{dop23}) is $\frac{{{{\left[ {{{\vec{N}}_\mathrm{AB}} \cdot \left( {{{\vec{v}}_\mathrm B} - {{\vec{v}}_\mathrm A}} \right)} \right]}^2}}}{{c{R_\mathrm{AB}}}}{T_{34}}$, which will achieve $3\times10^{-14}$. Hence, assuming that $T_{34}=100\ \mathrm{ns}$, the maximal Doppler residual errors caused by link 2 and link 3 are $3\times10^{-14}$. Since $\left( {\frac{{f_2^2}}{{f_1^2}} - 1} \right)\left( {\frac{{f_3^2}}{{f_3^2 - f_2^2}}} \right)$ in equation (\ref{m12f}) is approximately $-0.0053$, this error will affect gravitational redshift equation (\ref{tfc}) by a magnitude of $5\times10^{-16}$.

In \ref{sec:b}, we show the relevant approximations. We neglect the terms of $c^{-4}$, whose value are less than $10^{-17}$.

Above all, the maximal Doppler residual errors are $1.5\times10^{-16}$.

\subsubsection{Ionospheric residual errors}
\label{sec:4_1_2}
In the expansion of the ionosphere refractive index, the terms of $f^{-3}$ cannot be eliminated by our TFC method and have not been considered in the former sections. According to the study of \cite{Hoque2008}, the $f^{-3}$ terms of GNSS signals (L band) is approximately one in a hundred of $f^{-2}$ terms. By reckoning of our simulations, final errors caused by the $f^{-3}$ terms will be approximately $1\times10^{-15}$. Because of its variability and randomness, these errors will be effectively weakened to a considerably small value by averaging with a mass of data. The simulations in section \ref{sec:5} will demonstrate that this effect can be lower than $10^{-16}$.

\subsubsection{Relativistic residual errors}
\label{sec:4_1_3}
The calculation of relativistic effects will also cause residuals. There are differences among relativistic effects of links 1, 2 and 3; however, in our analysis, we consider it a constant value. The relativistic differences between link 2 and link 3 are much smaller than those between link 1 and link 2 (the time difference is much smaller), so we will only consider the differences between link 1 and link 2. Relativistic effects include gravitational redshift effect and transverse Doppler effect. Regardless of ground displacements such as the solid Earth tide, the gravitational redshift effect only varies with ISS, but the transverse Doppler effect varies with both ISS and ground station. Based on simulated data, the gravitational potential of ISS varies by $100 \ \mathrm{m^2/s^2}$ per second. However, $T_{23}$ is at most $1\ \mathrm{\mu s}$, so the gravitational redshift difference between link 1 and link 2 is approximately $10^{-20}$.

The transverse Doppler frequency shift is proportional to the square of velocity, and its differential expression is
\begin{equation}\label{dftrans}
\left| {d{f_\mathrm{trans}}} \right| = \frac{v}{{{c^2}}}dv = \frac{{va}}{{{c^2}}}dt
\end{equation}

Assuming that $T_{23}=1\ \mathrm{\mu s}$ and $T_{15}=2\ \mathrm{ms}$, the maximal transverse Doppler frequency shift errors caused by ISS and the ground station are $7.6\times10^{-19}$ and $3\times10^{-19}$, respectively. Thus, the maximal transverse Doppler frequency shift error is $1\times10^{-18}$.

\subsubsection{Tidal effects}
\label{sec:4_1_4}
With regard to ground displacements, the relativistic effects have other residuals. According to IERS convention 2010 \cite{Petit2010}, displacements of reference points are divided into three categories: (1) Tidal motions (mostly near diurnal and semidiurnal frequencies) and other accurately modeled displacements of reference markers (mostly at longer periods); (2) other displacements of reference markers including nontidal motions associated with the changing environmental loads; (3) displacements that affect the internal reference points in the observing instruments.

We are interested in the first two types. Tidal motions include solid, ocean and polar tides. The solid tide has the largest magnitude, which will be tens of centimeters \cite{Wahr1995,Petit2010}. Other displacements include nontidal mass redistributions in the atmosphere, oceans, sea-level variations, etc., but their amplitude is much smaller (at most several centimeters) \cite{Voigt2016}. Crustal deformation and plate motions also contribute to the parameters (${\vec r}_\mathrm B$, ${\vec v}_\mathrm B$, ${\vec a}_\mathrm B$), which can be predicted by the tectonic model NUVEL-1 NNR of \cite{Argus1991}. However, the numerical values of the kinematic vectors shows that the rates are only approximately several centimeters per year or tens of micrometers per day, which is 4 orders of magnitude smaller than the solid Earth tide ($\sim$ tens of centimeters per day). Therefore, we will take the solid Earth tide as the dominant contribution.

Based on the theory as proposed in IERS convention \cite{Voigt2016, Hartmann1995}, we estimated the equatorial displacement caused by solid tides, and the maximum is $0.32$ m. If a sine function is used to estimate the change in displacement with time, we assume that all tides are diurnal tides, and the effect of the tides on the velocity is $4.65\times10^{-5}$ m/s. According to those estimates, the magnitude of the position and velocity vectors affects the residual Doppler terms in (\ref{tfc}) and (\ref{tfc2}) by at most $6.76\times10^{-17}$, which is much less than the Doppler residuals analyzed in section \ref{sec:3_1}.

The solid Earth tide causes displacements and gravitational potential disturbance. The maximum tidal potentials caused by the moon and the sun are $4.41\ \mathrm{m^2/s^2}$ and $1.60\ \mathrm{m^2/s^2}$, respectively \cite{Hartmann1995}. With regard to Love numbers $h$, $l$ and $k$, the gravitational potential tide will be $3.74\ \mathrm{m^2/s^2}$, which is equivalent to a gravitational redshift of $4.2\times10^{-17}$, which is mainly caused by the semidiurnal tide.

Based on the above discussion in section \ref{sec:4_1}, the residual amount of each frequency shift is shown in table \ref{table1}. Among them, the Doppler frequency shift and ionosphere-related frequency shift are relatively large and can be manually eliminated if possible. The ionospheric part is difficult to manually eliminate, as demonstrated in further research (section \ref{sec:5}).

\begin{table*}
  \caption{Relative magnitudes and residual amount of each frequency shift in the TFC method}
  \footnotesize\rm
  \begin{tabular}{@{}l l l}
    \br
    Type & Magnitude & Residual \\
    \mr
    Doppler frequency shift & $10^{-5}\sim10^{-6}$ & $<1.5\times10^{-16}$ \\

    Ionospheric frequency shift along with its refraction effects & $10^{-10}\sim10^{-12}$ & Approximately $1\times10^{-15}$ \\

    Tropospheric frequency shift & $10^{-13}\sim10^{-14}$ & $<1\times10^{-18}$ \\

    Gravitational redshift  & Approximately $10^{-11}$ & - \\

    Tidal effects draw on gravitational redshift & $4.2\times10^{-17}$ & - \\

    Transverse Doppler frequency shift & $10^{-10}\sim10^{-11}$ & $<1\times10^{-18}$ \\

    Shapiro frequency shift & Approximately $10^{-14}$ & $<1\times10^{-18}$ \\
    \br
    \end{tabular}%
  \label{table1}%
\end{table*}%

\subsection{Test of gravitational redshift}
\label{sec:4_2}
In a static gravitational field, suppose that two atomic clocks are located at different positions. Then, we compare their frequencies by a certain frequency transfer method. Based on general relativity, gravitational redshift $\Delta \nu $ between clocks is proportional to their gravitational potential difference $\Delta U$ as
\begin{equation}\label{dnu}
\frac{{\Delta \nu }}{\nu } = \frac{{\Delta U}}{{{c^2}}}
\end{equation}

To test the gravitational redshift by a standard convention, parameter $\alpha$ is introduced via the following expression \cite{Will2014}
\begin{equation}\label{z}
 z = \frac{{\Delta \nu }}{\nu } = \left( {1 + \alpha } \right)\frac{{\Delta U}}{{{c^2}}}
\end{equation}
where $\alpha$ vanishes when Einstein equivalence principle (EEP) is valid. 

In our study, we develop a similar equation using another parameter $\beta$
\begin{equation}\label{z2}
z = \Delta {U_\mathrm m} = \left( {1 + \beta } \right)\Delta U
\end{equation}
where $\Delta U_\mathrm m$ is the measured gravitational potential difference by equations (\ref{tfc}) and (\ref{tfc2}), and $\Delta U$ is the standard gravitational potential difference developed by the Earth gravity field model. There are testing errors in both $\Delta U_\mathrm m$ and $\Delta U$, and the corresponding uncertainties should be calculated using the following equation 
\begin{equation}\label{u}
u = \sqrt {u_\mathrm{\Delta {U_m}}^2 + {{\left( {1 + \beta } \right)}^2}u_\mathrm{\Delta U}^2}
\end{equation}
where $u_\mathrm{\Delta {U_\mathrm m}}$ and $u_\mathrm{\Delta U}$ are the uncertainties of $\Delta U_\mathrm m$ and $\Delta U$, respectively.

\section{Simulation experiments and results}
\label{sec:5}
At present, there are no real data. To test our theory and formulations, we present simulation experiments. In these experiments, we use the data of the ISS real orbit, ionosphere, troposphere, calculated gravitational potential by the widely used gravity field model EGM2008 \cite{Pavlis2008}, solid Earth tide \cite{Hartmann1995}, and simulated clock data by a conventionally accepted stochastic noises model \cite{Allan1991,Galleani2003}.
\subsection{Simulation setup and experiments}
\label{sec:5_1}
In our simulations, we select the station Observatoire de Paris (OP) as the ground station with geographical parameters as shown in table \ref{table2}. In section \ref{sec:3_2}, we discussed that the magnitude of the tidal effect on gravitational redshift is approximately $10^{-17}$; nonetheless, in addition to the accuracy of the ACES program, we still added the tidal effect to the simulation experiment. When tidal effects are neglected, the relevant parameters in ECEF related to the ground station OP are considered constant, as shown in table \ref{table2}. Other parameter settings in our experiment are listed in table \ref{table3}.
    
\begin{table*}
  \centering
  \caption{Geographical parameters of OP}
    \footnotesize\rm
    \begin{tabular}{@{}l l l l l}
    \br
    Parameters & Latitude & Longitude & Height & Gravitational potential \\
    \mr
    Values & $48.836^{\circ}$ N & $2.336^{\circ}$ E & $124.2$ m & $62573855.538 \ \mathrm{m^2\cdot s^{-2}}$ \\
    \br
    \end{tabular}%
  \label{table2}%
\end{table*}%

\begin{table*}
  \centering
  \caption{Relevant parameters in the simulation experiment}
    \footnotesize\rm
    \begin{tabular}{@{}l l l l}
    \br
    Parameter & Value & Parameter & Value \\
    \mr
    Earth radius $R$ & $6378137$ m & Threshold of observation elevation & $15^{\circ}$ \\

    Earth flatness $e^2$ & $0.006694$ & Light speed in vacuum $c$ & $299792458$ m/s \\

    $T_{23}$   & $1\times10^{-6}$ s & Peak ionospheric height & $200$ km  \\

    $T_{34}$   & $1\times10^{-7}$ s & Peak electron density & $3\times10^{12} \ \mathrm{m^{-3}}$ \\

    Gravitational constant $GM$ & $3.9860\times10^{14}$ & Doodson's constant & $26277\ \mathrm{cm^2/s^2}$ \\
    \br
    \end{tabular}%
  \label{table3}%
\end{table*}%

\begin{figure}  
\centering  
\includegraphics[width=11cm]{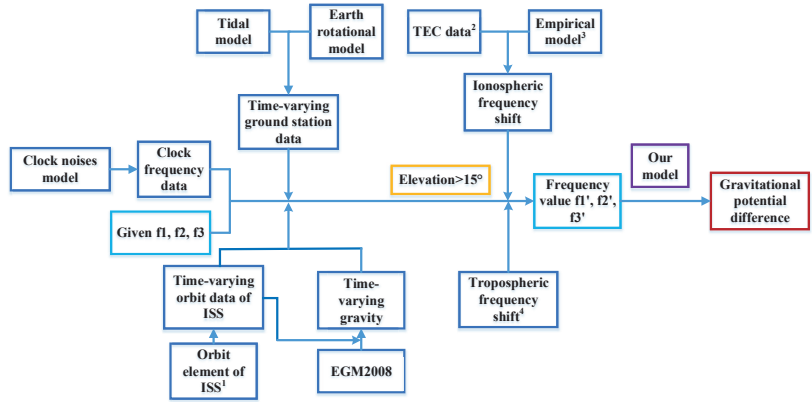}  
\caption{Diagram of the simulation experiments. The superscripts in red in this figure indicate that: 1. six orbit elements obtained from http://spaceflight.nasa.gov/; 2. TEC data obtained from ftp://cddis.gsfc.nasa.gov/; 3. Model of the refractive index varying with height; 4. Wet and dry ZTDs obtained from http://ggosatm.hg.tuwien.ac.at/.}
\label{fig3}  
\end{figure}

Figure \ref{fig3} shows the procedures of the simulation experiments. The observations in our simulation experiments are carrier frequency values of ACES, which are denoted as $f_1^{\prime}$, $f_2^{\prime}$, and $f_3^{\prime}$, as described in section \ref{sec:2}. We can use a combination of $f_1^{\prime}/f_1$, $f_2^{\prime}/f_2$, $f_3^{\prime}/f_3$ (as section \ref{sec:3} demonstrates) to calculate the gravitational potential difference between ISS and ground station and test the gravitational redshift.

To obtain the received frequency values ($f_1^{\prime}$, $f_2^{\prime}$, and $f_3^{\prime}$), the original emitted frequency values and various types of frequency shifts are required: (1) Originally emitted frequency values of $f_1$, $f_2$, and $f_3$, and clock noises (including devices noises); (2) frequency shifts including Doppler frequency shift, relativistic frequency shift (including second-order Doppler shift and gravitational redshift), atmospheric frequency shifts (including ionospheric and tropospheric parts), and tidal effects. To calculate these effects, we must have the position, velocity and acceleration information, which can be derived from the orbits of ISS and moving positions of the ground station. To select observable data, we set the condition that the observation elevation angle should be larger than $15^{\circ}$. 

First, we must solve the emitted frequency values, which are frequency series composed of the given frequencies $f_\mathrm{1}$, $f_\mathrm{2}$, $f_\mathrm{3}$ and clock noises. Based on the stochastic noise nature of the clocks, there are five types of clock noises: Random Walk FM (frequency modulation), Flicker FM, White FM, Flicker PM (phase modulation) and White PM \cite{Allan1991} with spectral densities of the types $f^{-2}$, $f^{-1}$, $f^{0}$, $f^{1}$ and $f^{2}$, respectively. 

\begin{figure}  
\centering  
\includegraphics[width=10cm]{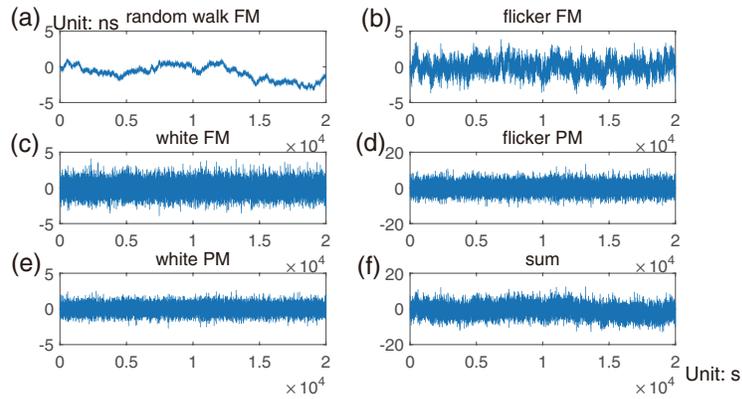}  
\caption{Sample of simulated clock noises. (a) Random walk FM series; (b) Flicker FM series; (c) White FM series; (d) Flicker PM series; (e) White PM series; (f) Sum of those five types of noises.}
\label{fig4}  
\end{figure}

Before starting our experiments, we provide samples of the simulated clock frequency noise series in figure \ref{fig4}. We simulated five pure types of noises with similar magnitudes (figures \ref{fig4}(a-e)) and their sum (figure \ref{fig4}f). The data length is 20\,000. Figure \ref{fig4}a shows a strong trend, figure \ref{fig4}b shows little trend and some periodic patterns, and figures \ref{fig4}(c-e) show irregular patterns. White and random walk noises are the simple types \cite{Galleani2003}. Flicker noises were calculated by the AR (autoregressive) model \cite{Kasdin1995}. 

\begin{figure}  
\centering  
\includegraphics[width=8cm]{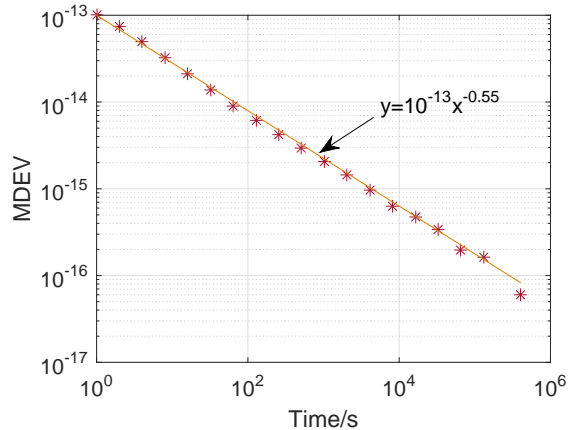}  
\caption{Modified Allan Deviation (MDEV) of the simulated clock errors. The points refer to the calculated MDEV data, and the line is fitted from the points.}
\label{fig5}  
\end{figure}

In simulations, we took White FM as the dominant part because Allan deviation performance of PHARAO is very similar to that of pure white FM, and we simulated 864\,000 s of clock data with sample intervals of 1 s. The modified Allan deviation (MDEV, \cite{Allan1991}) of our simulated data is shown in figure \ref{fig5}, and its performance is similar with previous studies \cite{Cacciapuoti2009,Cacciapuoti2017}. Figure \ref{fig5} shows a potential of long-term stability of $10^{-16}$. With these clock data, we succeeded in the first step: the emitted frequency values were generated. 

In the experiment, to simulate the orbit, we use the daily orbit elements to calculate. The TEC (Total Electron Content) data are interpolated from the grid data, which are derived from CDDIS (The Crustal Dynamics Data Information System). Using the TEC data and time-varying rate, we can simulate the ionospheric frequency shift. The refractive frequency shift caused by the ionosphere is calculated by the refractive angle, which is integrated from the layered structure adopted model \cite{Hajj1998} of ionosphere. The troposphere is calculated by the zenith tropospheric delay (ZTD) of the dry and wet components, and the projection function adopted the Vienna Mapping Function (VMF1) mode. The refractive frequency shift caused by the ionosphere is not calculated because this part is independent of the carrier frequency and easy to eliminate. We adopted EGM2008 \cite{Pavlis2008} for gravitational potential models. We considered the effect of the tidal effect on the ground gravitational potential because this is the main source of residual error in gravitational redshift. For this part, we first calculated the Earth tide generated potential from parameters of periodic tides \cite{Hartmann1995} using the method of harmonic analysis. Then, we considered Love numbers and solved the gravitational potential tide. Combining with the analysis in section \ref{sec:4}, we did not consider the indirect tidal effect on the coordinates variations of the ground station because this indirect tide is far from other residual errors. 

\begin{figure}  
\centering  
\includegraphics[width=9cm]{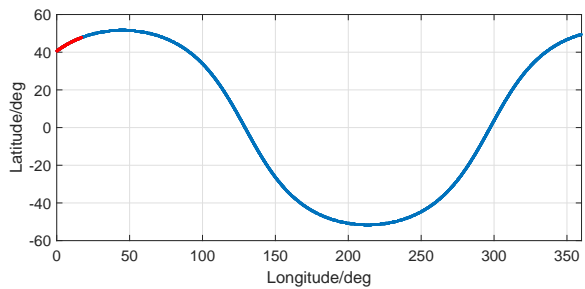}  
\caption{ISS’s track of the subsatellite point. The red part of the curve denotes the position of ISS that can be directly observed, and the blue part of the curve denotes the trajectory of an entire circle.}
\label{fig6}  
\end{figure}

ISS is flying in an orbit with an inclination of $51.6^{\circ}$ and a period of 5400 s, and its track of the subsatellite point is shown in figure \ref{fig6}. In this figure, the red part is where we can observe. Because ISS is flying in a low orbit with a large velocity, each pass over the ground station will last approximately 600 seconds at most \cite{Meynadier2018}; for our elevation threshold of $15^{\circ}$, we can only observe approximately 300 s per pass.

\subsection{Results and accuracy level of the test}
\label{sec:5_2}
\begin{figure}  
\centering  
\includegraphics[width=9cm]{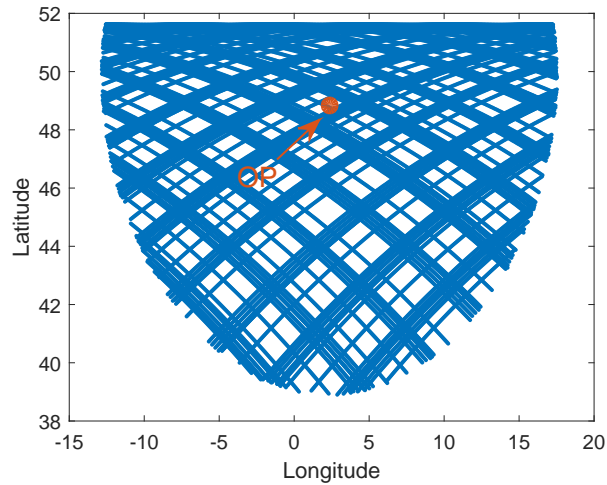}  
\caption{Track of the subsatellite point of all passes of ISS in 29 days of observation.}
\label{fig7}  
\end{figure}

For our experiment, we had 29-day simulated observations, and ISS flew above OP 130 times in total. As shown in figure \ref{fig7}, we drew a subsatellite point trajectory with the station OP at the center. Because the inclination of ISS is $51.6^{\circ}$, the maximal latitude of the subsatellite point is that value.

After analyzing the orbits of ISS, we selected the time when we could observe with elevations greater than $15^{\circ}$. Based on the position information of ISS and OP, downloaded atmospheric parameters, gravitational quantities and various models, we calculated all types of frequency shifts as shown in figure \ref{fig3}, whose magnitudes are shown in table \ref{table1}. The result shows that Doppler frequency shift is the dominant part, while Shapiro frequency shift is the smallest part. Refraction effects cannot be neglected because they have larger magnitude than the ionospheric frequency shift and tropospheric frequency shift. In the experiment, although the residuals of some errors (especially the higher-order term of the ionosphere) are greater than $10^{-16}$, due to the randomness of the errors, simulation experiments show that after a long-term average, they can be less than $10^{-16}$.

\begin{figure}  
\centering  
\includegraphics[width=9cm]{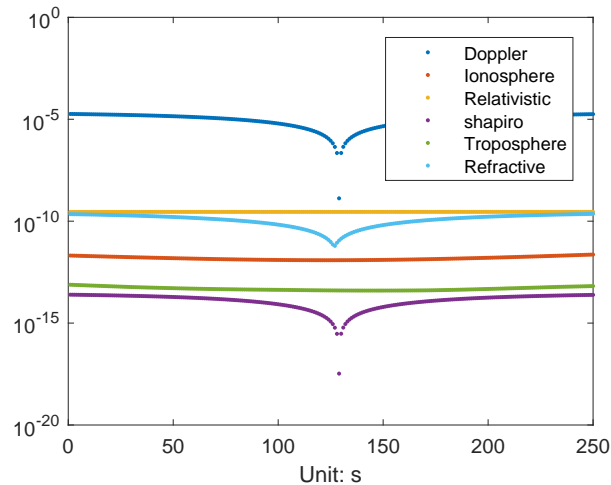}  
\caption{Each frequency shift in a one-way transfer.}
\label{fig8}  
\end{figure}

\begin{figure}  
\centering  
\includegraphics[width=9cm]{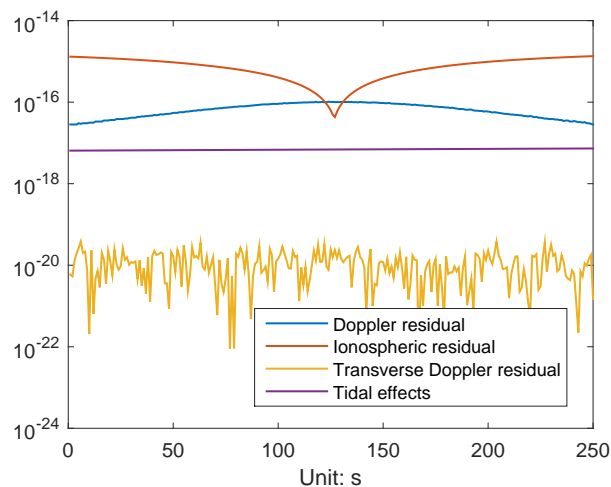}  
\caption{Residual errors of various frequency shifts.}
\label{fig9}  
\end{figure}

Furthermore, we analyzed each frequency shift in a one-way transfer (link 3, figure \ref{fig8}) and the residual of various frequency shifts after applying the TFC method (figure \ref{fig9}). For the time length, we only show one pass. Figure \ref{fig8} is consistent with table \ref{table1}, where the relativistic frequency shift appears unchanged because the gravitational potential and velocity norm of ISS slowly change with time. There appears to be singularities in figure \ref{fig8}, but those are not real. When ISS was exactly on top of the station, the velocity became vertical to the line of sight (LoS), which made Doppler and Shapiro effects extremely small and resemble a singularity.

The TFC method can largely eliminate various errors, but the analysis in section \ref{sec:4_1} shows that residual errors remain. According to table \ref{table1}, we calculated the Doppler residuals, ionospheric-related residuals (including ionospheric frequency shift and ionospheric refraction), transverse Doppler residuals, and tidal effects. The Doppler residual is caused by the Doppler difference between link 2 and link 3. The ionospheric-related residual is caused by the higher-order ionospheric term. The transverse Doppler residual is caused by the change in velocity. Figure \ref{fig9} shows that among the residuals of various frequency shifts, the largest component is the ionospheric residual, which can reach a maximum of $1\times10^{-15}$ and is often at the level of $10^{-16}$; the second largest component is the Doppler residual, whose maximum is $1\times10^{-16}$; the magnitudes of other frequency shifts are much smaller than $10^{-16}$, so they are negligible. These two frequency shifts should be corrected in the TFC method model, but simulation experiments show that due to the variability of these two frequency shifts, they can be eliminated below $10^{-16}$ after a long-term average, so they can be ignored. Thus, the TFC method model can indeed eliminate the largest Doppler frequency shift. The magnitudes of various residual errors are consistent with the estimates in table \ref{table1}.

In addition to the discussed systematic residual errors, random errors will be caused by the parameters of the ground station and space station in the results. We will later analyze the impact of these errors.

\begin{figure}  
\centering
\subfigure{
\begin{minipage}[t]{0.5\linewidth}
\centering
\includegraphics[width=7cm]{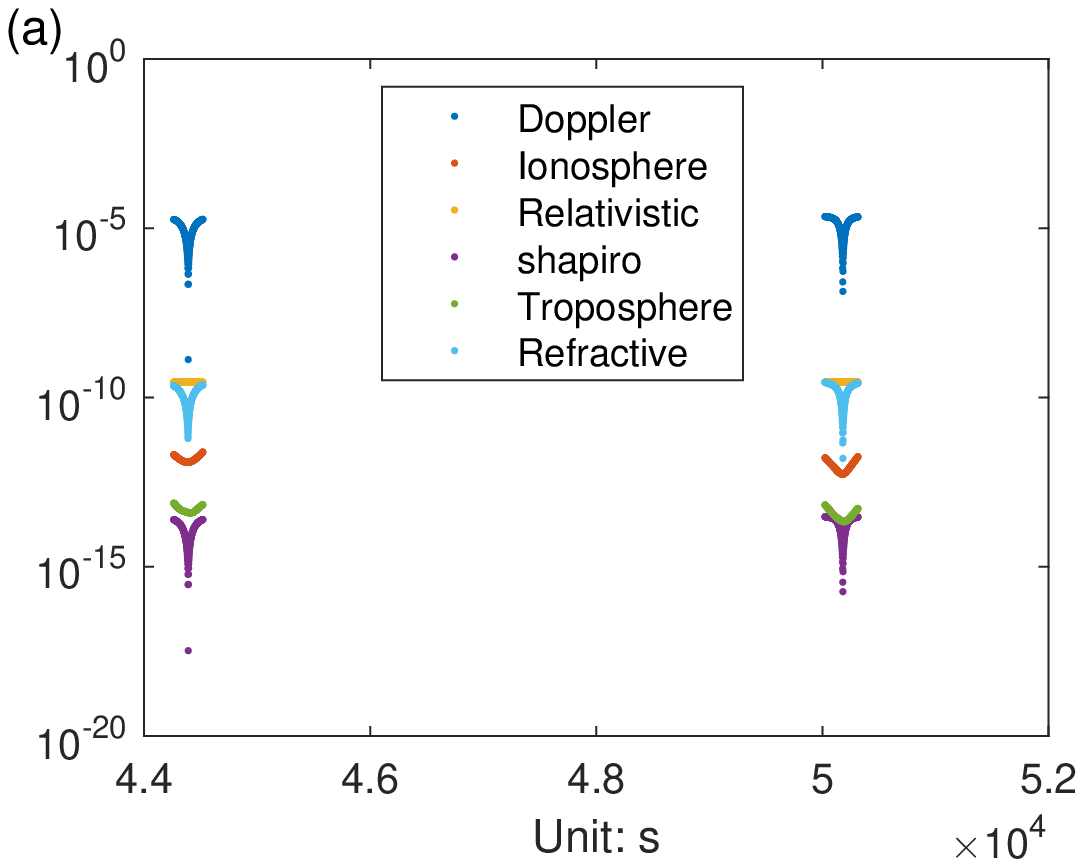}
\end{minipage}%
}%
\subfigure{
\begin{minipage}[t]{0.5\linewidth}
\centering
\includegraphics[width=7cm]{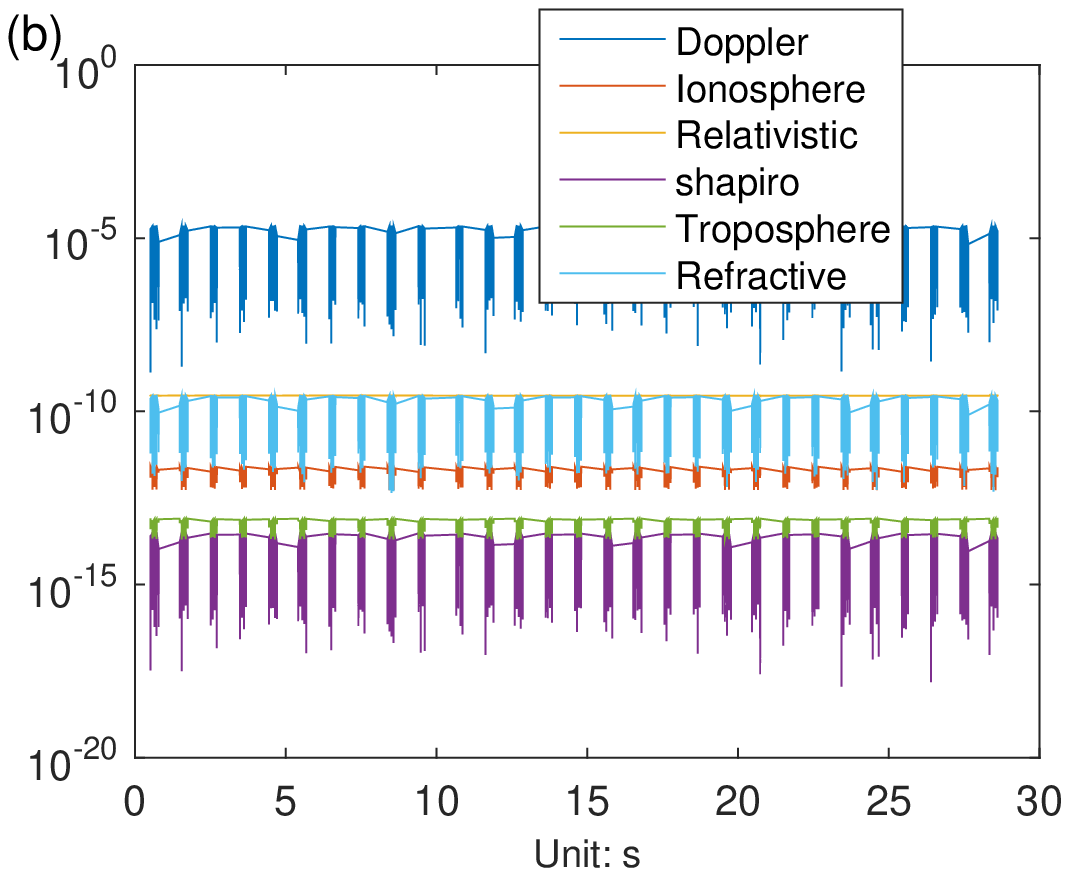}
\end{minipage}%
}%
\caption{(a) Various frequency shifts in a one-way link of two adjacent epochs; (b) Various frequency shifts in a one-way link in the 29-day observation.}
\label{fig10}  
\end{figure}

Figure \ref{fig8} and figure \ref{fig9} only show the frequency shift during a single flight of the space station, and figure \ref{fig10} shows the data for a longer time period. We define the observations of one single pass of ISS as an epoch. Figure \ref{fig10}(a) shows 2 closed epochs, and figure \ref{fig10}(b) shows the entire data (130 epochs). Time intervals between two epochs are very large and much larger than the time duration of one epoch itself. Thus, we can solve these data epoch by epoch, obtain the averaged results, and evaluate the results of each epoch. 

\begin{figure}  
\centering  
\includegraphics[width=9cm]{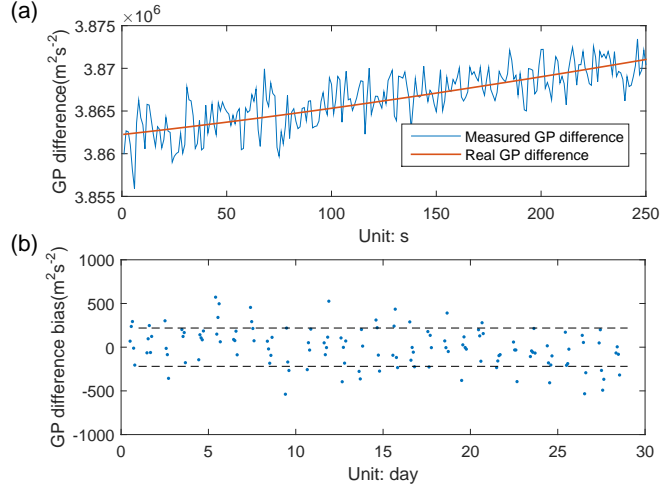}  
\caption{(a) Various frequency shifts in a one-way link of two adjacent epochs; (b) Various frequency shifts in a one-way link in the 29-day observation.}
\label{fig11}  
\end{figure}

Finally, we obtained the received frequency values ($f_1^{\prime}$, $f_2^{\prime}$ and $f_3^{\prime}$). Using our TFC model as section \ref{sec:3} proposed, we can obtain the gravitational potential differences (blue line) and compared them with the real differences calculated by EGM2008 in figure \ref{fig11}(a). The results are obtained by only one single epoch, and the dotted line in the figure is the range of the error in the $1-\sigma$ criterion. Figure \ref{fig11}(b) shows the gravitational potential differences after averaging epoch by epoch. The results show that the precision is not good for data sampled per second (figure \ref{fig11}(a)), which is equivalent to 250 m in height, which is within our expectation. However, after the epoch-by-epoch averaging, we obtained a gravitational potential difference bias series. The precision of the gravitational potential differences has improved to approximately $218\ \mathrm{m^2/s^2}$ (approximately equivalent to an elevation of 22.2 m), as shown in figure \ref{fig11}(b). After averaging the entire data set, we obtain a gravitational potential difference bias of $4.0\pm18.6\ \mathrm{m^2/s^2}$. Compared with the real averaged gravitational potential difference of $3.8340\times10^6 \ \mathrm{m^2/s^2}$, our testing level achieves $(1.04\pm4.85)\times10^{-6}$. Supposing that the gravitational potential difference calculated by EGM2008 has an uncertainty of $3\ \mathrm{m^2/s^2}$, the testing uncertainty achieves ${{\sqrt {{{18.6}^2}+{3^2}} } \mathord{\left/
 {\vphantom {{\sqrt {{{18.6}^2}+{3^2}} } {\left( {3.8340 \times {{10}^6}} \right)}}} \right.
 \kern-\nulldelimiterspace} {\left( {3.8340 \times {{10}^6}} \right)}} \approx 4.91 \times {10^{ - 6}}$; therefore, our testing level is $(1.04\pm4.91)\times10^{-6}$. Here, we only used 29 days of observations. With longer-period experiments, the results will be better.

\begin{table}
  \centering
  \caption{Gravitational potential difference and its testing level}
    \footnotesize\rm
    \begin{tabular}{@{}l l l}
    \br
    Parameters & Value & Uncertainty \\
    \mr
    Model & $3834161.3\ \mathrm{m^2\cdot s^{-2}}$ & $3\ \mathrm{m^2\cdot s^{-2}}$ \\
    
    Observation & $3834165.3\ \mathrm{m^2\cdot s^{-2}}$ & $18.6\ \mathrm{m^2\cdot s^{-2}}$ \\

    Testing level & $1.04\times10^{-6}$ & $4.91\times10^{-6}$ \\
    \br
    \end{tabular}%
  \label{table4}%
\end{table}%

\subsection{Accuracy requirements of relevant parameters}
\label{sec:5_3}
For the calculation of our TFC model, we examine the requirements of the parameter precision (table \ref{table5}). To obtain results with high precision, we need parameters to satisfy these demands. ${\vec r}_\mathrm A$ and ${\vec r}_\mathrm B$ are the positions of ISS and the ground station, respectively; ${\vec v}_\mathrm A$ and ${\vec v}_\mathrm B$ are their velocities; ${\vec a}_\mathrm A$ and ${\vec a}_\mathrm B$ are their accelerations;  $T_{23}$ is the time offset of link 1 and link 2. All ground station parameters here can be accurately obtained based on the coordinates and Earth rotation model and do not need to be discussed in detail. Thus, we must only study the parameters related to the space station.

These parameters are demanded for the calculation of equation (\ref{tfc}) and (\ref{tfc2}); thus, we derived the total differential of these equations and analyzed the differential relationship between parameters (${\vec r}_\mathrm A$, ${\vec v}_\mathrm A$, ${\vec a}_\mathrm A$ and $T_{23}$) and gravitational potential difference. The measurement noises are stochastic, so after averaging, the noises will be largely weakened. Due to this principle, based on the MDEV curve of ACES clocks, we set a target precision of $10^{-13}$ at a sampling rate of 1 second.

According to table \ref{table5}, the requirements of orbit parameters and acceleration parameters are easy to satisfy, and $T_{23}$ is easy to control at that level. Only the precision requirement of the velocity is high and must be carefully solved. It is easy for ISS to satisfy these demands. In our parameter analysis, for each analysis of one parameter, we suppose that the other parameters are constant. In this case, we may clearly examine the effect of the error of each parameter on the results. In the real environment, there are stochastic errors in all parameters. Table \ref{table5} shows that only the velocity is the principal error source if all errors exist. Therefore, our parameter analysis scheme is feasible.
\begin{table}
  \centering
  \caption{Minimal required accuracy of parameters calculated based on simulation data; in the case, the accuracy requirement for gravitational redshift is $1\times10^{-16}$}
    \footnotesize\rm
    \begin{tabular}{@{}l l l l}
    \br
    Parameters & Demand along the rail & Horizontal demand & Radial demand \\
    \mr
    ${\vec r}_\mathrm A$    & $240\ \mathrm m$ & $689\ \mathrm m$ & $465\ \mathrm m$ \\

    ${\vec v}_\mathrm A$    & $1.23\ \mathrm{m/s}$ & $3.26\ \mathrm{m/s}$ & $2.48\ \mathrm{m/s}$ \\

    ${\vec a}_\mathrm A$    & $69.6\ \mathrm{m/s^2}$ & $80.0\ \mathrm{m/s^2}$ & $65.3\ \mathrm{m/s^2}$ \\

    $T_{23}$   & \multicolumn{3}{c}{$5.5\times10^{-7}$ s} \\
    \br
    \end{tabular}%
  \label{table5}%
\end{table}%

\section{Conclusion}
\label{sec:6}
In this study, first we proposed a new one-way frequency transfer model based on the frequency transfer theory to order $c^{-3}$ \cite{Blanchet2001} and Doppler shift considering the refractive effect \cite{Bennett1968}. Our model is established in free space and has an accuracy of $c^{-3}$. Then, we proposed the TFC method to derive the gravitational redshift based on three frequency links of ACES. Unlike other combination methods of ACES \cite{Duchayne2009,Meynadier2018}, our model addresses the frequency comparison instead of time comparison. This combination method can largely eliminate Doppler frequency shift, atmospheric effects, etc. We also calculated the residual errors and test whether they could be eliminated. Furthermore, we designed simulation experiments to authenticate our model and analyzed the demanded parameters. The simulation results show that our testing level can achieve $(1.04\pm4.91)\times10^{-6}$.

Equipped with ACES atomic clocks (SHM and PHARAO) with a frequency stability of $(1-3)\times10^{-16}$, our study shows that the test of gravitational redshift can achieve an accuracy level of $2\times10^{-6}$ if longer data (say much longer data than 29 days) are obtained, which is one and a half orders higher than the accuracy level given by \cite{Vessot1980}. This study provides a new approach to test the gravitational redshift and benefits the brand-new field of relativistic geodesy.

ACES limits its testing precision at the $10^{-6}$ level due to the limitation of the accuracy of the atomic clock it carries. To achieve higher scientific goals, more precise space atomic clocks and time-frequency comparison links are required. The Chinese Space Station (CSS) provides this opportunity. Since the CSS will carry optic-atomic clocks on the order of $10^{-18}$, which is one and a half orders of magnitude higher than the cold atomic cesium clocks of ACES. It is expected to achieve 10-cm to 5-cm level gravitational potential measurements and can test general relativity at a level of $10^{-7}$.

However, due to the higher accuracy of the clock on the CSS, there are differences in model and actual measurement. In terms of the model, it must consider the time-frequency comparison model under the relativistic framework of $c^{-4}$ accuracy, and some of the originally negligible residual items in section \ref{sec:4_1} must be carefully considered. In addition, because the magnitude of the high-order ionospheric term is obvious and difficult to eliminate, one may consider a potential solution to generalize the TFC method to a four-frequency combination and remove the high-order ionospheric terms. If the hardware conditions cannot satisfy the requirements of the quad-band link, a more in-depth study of the ionospheric frequency shift is required. If the accuracy of the model must reach the level of $10^{-18}$, it is also necessary to consider the tidal effect.

In terms of actual measurement, due to the higher accuracy of gravitational potential, according to the model, there are higher-accuracy requirements in measuring the position and speed of the space station, which introduces further requirements on the positioning system of the space station and hardware facilities for time measurement. If the accuracy requirement is increased by one and a half orders of magnitude, the velocity of space station should achieve the centimeter accuracy level. In summary, the formulation of the TFC method proposed in this study is also feasible for the CSS plan.

\ack This study is supported by NSFCs (grant Nos. 41631072, 41721003, 41574007, 41874023, 41804012), and Natural Science Foundation of Hubei Province (grant No. 2019CFB611). \\

\appendix
\section{}
\label{sec:a}
In our modeling, we focus on the down link from A to B, but since the optical path is reversible, we draw figure \ref{fig1} as if it is the link from B to A. Assuming that our Earth is a sphere, and the atmosphere is composed of many thin spherical shells, in each of which the medium is homogeneous. Based on Bouguer’s law (Snell’s law in spherical surface) \cite{Croft1968,Hoque2008}, the following equation holds
\begin{equation}\label{a}
a = nr\sin i = const
\end{equation}
where $n$ is the refractive index, and $i$ is the zenith angle at each layer.

For electromagnetic waves that propagate in a medium with refractive index $n$, the differential bending angle $d{\bm{\alpha }}$  that accrue on the ray path over a differential arc length $ds$ is given by \cite{Melbourne1994}
\begin{equation}\label{dalpha}
d{\bm{\alpha }} = {n^{ - 1}}\left( {{\vec{T}} \times \nabla n} \right)ds
\end{equation}
where $\vec T$ is its unit tangent vector defined by ${\vec T}={\vec k}/k$, ${\vec k}(r)$ is the vector wavenumber of the ray at point $r$, and $\Delta n$ is the gradient vector of the refraction index. Here, differential bending angle $d{\bm{\alpha }}$ is written in vector form and related to the value and direction. 

The refraction index in the ionosphere is displayed by equation (\ref{n})
\begin{equation}\label{n2}
n = 1 - 40.3\frac{{{n_\mathrm e}}}{{{f^2}}}
\end{equation}
We suppose that the refractive index only changes in the vertical direction. At the lower middle part of the ionosphere layer, electron density ${n_\mathrm e}$ is at the peak \cite{Hajj1998}. If we integrate equation (\ref{dalpha}) from the peak height to the ISS height, the gradient of the refractive index can be expressed as a vertical downward vector form, and the integration is
\begin{equation}\label{alpha}
{\bm{\alpha }} = \int_P {{n^{ - 1}}\left( {{\vec{T}} \times \nabla n} \right)ds}  = \int_\mathrm{{h_\mathrm{\max }}}^{{h_\mathrm{ISS}}} {{n^{ - 1}}\frac{{dn}}{{dh}}\frac{{\sin i}}{{\cos i}}{{\vec{e}}_\mathrm g}dh}
\end{equation}
where ${{\vec{e}}_\mathrm g}$ is the unit vector whose direction is defined by ${\vec{T}} \times \nabla n$. According to equation (\ref{n2}), ${{\partial n} \mathord{\left/
 {\vphantom {{\partial n} {\partial h}}} \right.
 \kern-\nulldelimiterspace} {\partial h}}$ is a positive value, and we use
\begin{eqnarray}\label{alpha2}
\alpha  &= \int_0^{{h_\mathrm{ISS}}} {{n^{ - 1}}\frac{{dn}}{{dh}}\frac{{\sin i}}{{\cos i}}dh}  = \int_0^{{h_\mathrm{ISS}}} {{n^{ - 1}}\frac{{dn}}{{dh}}\frac{{\frac{a}{{n\left( {h + {R_\mathrm E}} \right)}}}}{{\sqrt {1 - {{\left( {\frac{a}{{n\left( {h + {R_\mathrm E}} \right)}}} \right)}^2}} }}dh} \nonumber\\
& = \int_0^{{h_\mathrm{ISS}}} {{n^{ - 1}}\frac{{dn}}{{dh}}\frac{a}{{\sqrt {{n^2}{{\left( {h + {R_\mathrm E}} \right)}^2} - {a^2}} }}dh} 
\end{eqnarray}
where $R_\mathrm E$ is the Earth’s average radius. 

For the links of ACES, we consider the ionospheric part (discussed in section \ref{sec:2}), and we let \cite{Hajj1998}
\begin{equation}\label{neh}
{n_\mathrm e}(h) = \cases{
{N_\mathrm{\max }}\exp \left( { - \frac{{h - {h_\mathrm{\max }}}}{H}} \right) & $h > {h_\mathrm{\max }}$\\
0 & otherwise\\
}
\end{equation}
where $h_\mathrm{max}$ and $N_\mathrm{max}$ correspond to the peak height and peak density, respectively; $H$ is the free electron density scale height.
Considering (\ref{alpha}) and (\ref{alpha2}), we have the following for the ionosphere
\begin{equation}\label{dndh}
\frac{{dn}}{{dh}} = \frac{{40.3{N_\mathrm{\max }}}}{{H{f^2}}}\exp \left( { - \frac{{h - {h_\mathrm{\max }}}}{H}} \right)
\end{equation}
Substituting (\ref{dndh}) into (\ref{alpha2}) and taking $n=1$ for approximation, equation (\ref{alpha2}) gives
\begin{equation}\label{alphaion}
{\alpha _\mathrm{ion}} = \frac{{40.3a{N_\mathrm{\max }}}}{{H{f^2}}}I
\end{equation}
where integration $I$ is
\begin{equation}\label{i}
I = \int_{{h_\mathrm{\max }}}^{{h_\mathrm{ISS}}} {\frac{1}{{\sqrt {{n^2}{{\left( {h + {R_\mathrm E}} \right)}^2} - {a^2}} }}\exp \left( { - \frac{{h - {h_\mathrm{\max }}}}{H}} \right)dh}
\end{equation}
It is weekly correlated to carrier frequency $f$, and the tropospheric part can be determined in this manner. For the total bending angle, we have
\begin{equation}\label{alphasum}
\alpha  = {\alpha _\mathrm{ion}} + {\alpha _\mathrm{trop}}
\end{equation}

Based on the reversible principle of the optical path, for three signals of ACES, the only difference is carrier frequency $f$. Moreover, we have $\alpha_\mathrm{ion}\sim1/f^2$. 

To calculate the effect of refraction on the Doppler effect, we must separately calculate the deflection angles at A and B. If we define $\delta_\mathrm A$ and $\delta_\mathrm B$ to be the deflection angle between the direction of the ray and the direct connection of A and B, which implies that
\begin{equation}\label{alphasum2}
\alpha  = {\delta _\mathrm A} + {\delta _\mathrm B}
\end{equation}
As figure \ref{fig1} shows, we have
\begin{eqnarray}\label{deltasum}
{\delta _\mathrm A} = {\beta _\mathrm A} - {\theta _\mathrm A}\nonumber\\
{\delta _\mathrm B} = {\theta _\mathrm B} - {\beta _\mathrm B}
\end{eqnarray}

According to equations (\ref{a}) and (\ref{deltasum}), we have
\begin{equation}\label{equaab}
{n_\mathrm A}{r_\mathrm A}\left( {\sin {\theta _\mathrm A} + {\delta _\mathrm A}\cos {\theta _\mathrm A}} \right) = {n_\mathrm B}{r_\mathrm B}\left( {\sin {\theta _\mathrm B} - {\delta _\mathrm B}\cos {\theta _\mathrm B}} \right)
\end{equation}
For equations (\ref{alphasum}), (\ref{alphasum2}) and (\ref{equaab}), $\delta_\mathrm A$ and $\delta_\mathrm B$ can be solved by the expression of $\alpha$
\begin{eqnarray}\label{deltasum2}
{\delta _\mathrm A} =  - \frac{{{n_\mathrm A}{r_\mathrm A}\sin {\theta _\mathrm A} - {n_\mathrm B}{r_\mathrm B}\sin {\theta _\mathrm B} + \alpha {n_\mathrm B}{r_\mathrm B}\cos {\theta _\mathrm B}}}{{{n_\mathrm A}{r_\mathrm A}\cos {\theta _\mathrm A} - {n_\mathrm B}{r_\mathrm B}\cos {\theta _\mathrm B}}}\nonumber\\
{\delta _\mathrm B} = \frac{{{n_\mathrm A}{r_\mathrm A}\sin {\theta _\mathrm A} - {n_\mathrm B}{r_\mathrm B}\sin {\theta _\mathrm B} + \alpha {n_\mathrm A}{r_\mathrm A}\cos {\theta _\mathrm A}}}{{{n_\mathrm A}{r_\mathrm A}\cos {\theta _\mathrm A} - {n_\mathrm B}{r_\mathrm B}\cos {\theta _\mathrm B}}}
\end{eqnarray}

From equation (\ref{deltasum2}), $n_\mathrm A$ and $\alpha$ are relevant with carrier frequency $f$ (ISS is in the ionosphere layer, the term $n_\mathrm A-1$ is proportional to $f^{-2}$, and the ground station is in the troposphere layer), and the term ${n_\mathrm A}{r_\mathrm A}\cos {\theta _\mathrm A} - {n_\mathrm B}{r_\mathrm B}\cos {\theta _\mathrm B}$ is nearly irrelevant to carrier frequency $f$ because the term $n_\mathrm A-1$ is small compared to the entire term. Thus, equation (\ref{deltasum2}) can be divided into two parts: the first part is irrelevant to carrier frequency $f$, and the second part is proportional to $f^{-2}$.
\begin{eqnarray}\label{deltasum3}
{\delta _\mathrm A} = \delta _\mathrm A^0 + \delta _\mathrm A^\mathrm{ion} =  - \frac{{{r_\mathrm A}\sin {\theta _\mathrm A} - {n_\mathrm B}{r_\mathrm B}\sin {\theta _\mathrm B} + {\alpha _\mathrm{trop}}{n_\mathrm B}{r_\mathrm B}\cos {\theta _\mathrm B}}}{{{n_\mathrm A}{r_\mathrm A}\cos {\theta _\mathrm A} - {n_\mathrm B}{r_\mathrm B}\cos {\theta _\mathrm B}}} \nonumber\\
\qquad - \frac{{\left( {{n_\mathrm A} - 1} \right){r_\mathrm A}\sin {\theta _\mathrm A} + {\alpha _\mathrm{ion}}{n_\mathrm B}{r_\mathrm B}\cos {\theta _\mathrm B}}}{{{n_\mathrm A}{r_\mathrm A}\cos {\theta _\mathrm A} - {n_\mathrm B}{r_\mathrm B}\cos {\theta _\mathrm B}}}\nonumber\\
{\delta _\mathrm B} = \delta _\mathrm B^0 + \delta _\mathrm B^\mathrm{ion} = \frac{{{r_\mathrm A}\sin {\theta _\mathrm A} - {n_\mathrm B}{r_\mathrm B}\sin {\theta _\mathrm B} + {\alpha _\mathrm{trop}}{n_\mathrm A}{r_\mathrm A}\cos {\theta _\mathrm A}}}{{{n_\mathrm A}{r_\mathrm A}\cos {\theta _\mathrm A} - {n_\mathrm B}{r_\mathrm B}\cos {\theta _\mathrm B}}} \nonumber\\
\qquad + \frac{{\left( {{n_\mathrm A} - 1} \right){r_\mathrm A}\sin {\theta _\mathrm A} + {\alpha _\mathrm{ion}}{n_\mathrm A}{r_\mathrm A}\cos {\theta _\mathrm A}}}{{{n_\mathrm A}{r_\mathrm A}\cos {\theta _\mathrm A} - {n_\mathrm B}{r_\mathrm B}\cos {\theta _\mathrm B}}}
\end{eqnarray}

To calculate ${\vec{T}} \cdot {\vec{v}}$ in equation (\ref{adopbar2}), we must project these two vectors to the same plane. We set point B to be the origin, direction OB to be the y-axis, and the tangent of the Earth surface to be the x-axis, and we obey the right-hand rule to define the z-axis. This coordinate system is defined as the Local Link Coordinates System (LLCS). We can transform vectors from ECEF to LLCS and select the x and y components in LLCS, which are the projected coordinates. In the xoy plane of LLCS, vectors ${\vec T}_\mathrm A$ and ${\vec T}_\mathrm B$ can be expressed as
\begin{eqnarray}\label{tatb}
&{{\vec{T}}_\mathrm A} = \left[ {\cos \left( {{\gamma _\mathrm B} - {\delta _\mathrm A}} \right),\sin \left( {{\gamma _\mathrm B} - {\delta _\mathrm A}} \right)} \right]\nonumber\\
&{{\vec{T}}_\mathrm B} = \left[ {\cos \left( {{\gamma _\mathrm B} + {\delta _\mathrm B}} \right),\sin \left( {{\gamma _\mathrm B} + {\delta _\mathrm B}} \right)} \right]
\end{eqnarray}

The velocity vectors (of both ISS and ground station) can be transformed by
\begin{eqnarray}\label{vllcsa}
&{{\vec{v}}_\mathrm{LLCS}} = {{\vec{A}}^{ - 1}}{\vec{v}}\nonumber\\
&{\vec{A}} = \left[ {\begin{array}{*{20}{c}}
{{{\vec{e}}_x}}&{{{\vec{e}}_y}}&{{{\vec{e}}_z}}
\end{array}} \right]
\end{eqnarray}

Here, $\vec{A}$ is the transition matrix, and ${\vec{e}}_x$, ${\vec{e}}_y$, and ${\vec{e}}_z$ can be solved by
\begin{eqnarray}\label{en}
{{\vec e}_x} &= \frac{{{{\vec n}_x}\left| {{{\vec n}_x} \cdot {{\vec r}_\mathrm{OA}}} \right|}}{{{{\vec n}_x} \cdot {{\vec r}_\mathrm{OA}}\left| {{{\vec n}_x}} \right|}}\nonumber\\
{{\vec e}_y} &= \frac{{{{\vec r}_\mathrm{OB}}}}{{\left| {{{\vec r}_\mathrm{OB}}} \right|}}\nonumber\\
{{\vec e}_z} &= {{\vec e}_x} \cdot {{\vec e}_y}\nonumber\\
{{\vec n}_x} &= {{\vec r}_\mathrm{OA}} - \frac{{{{\vec r}_\mathrm{OA}} \cdot {{\vec r}_\mathrm{OB}}}}{{{{\left| {{{\vec r}_\mathrm{OB}}} \right|}^2}}}{{\vec r}_\mathrm{OB}}
\end{eqnarray}
where ${{\vec r}_\mathrm{OA}}$ and ${{\vec r}_\mathrm{OB}}$ are vectors of OA and OB in ECEF.

We select the x and y components of ${\vec v}_\mathrm A$ and ${\vec v}_\mathrm B$ after coordinate transferring to be $v_\mathrm{Ax}$, $v_\mathrm{Ay}$, $v_\mathrm{Bx}$ and $v_\mathrm{By}$. Considering equations (\ref{inteindex}) and (\ref{tatb}), the first term of equation (\ref{adopbar2}) is expressed as
\begin{eqnarray}\label{ntv}
\frac{{1 - \frac{{{n_\mathrm B}{{\vec{T}}_\mathrm B} \cdot {{\vec{v}}_\mathrm B}}}{c}}}{{1 - \frac{{{n_\mathrm A}{{\vec{T}}_\mathrm A} \cdot {{\vec{v}}_\mathrm A}}}{c}}} = \frac{{1 - \frac{{{{\vec{N}}_\mathrm{AB}} \cdot {{\vec{v}}_\mathrm B}}}{c} - \frac{{\left( {{{\vec{T}}_\mathrm B} - {{\vec{N}}_\mathrm{AB}}} \right) \cdot {{\vec{v}}_\mathrm B}}}{c} - \frac{{\left( {{n_\mathrm B} - 1} \right){{\vec{T}}_\mathrm B} \cdot {{\vec{v}}_\mathrm B}}}{c}}}{{1 - \frac{{{{\vec{N}}_\mathrm{AB}} \cdot {{\vec{v}}_\mathrm A}}}{c} - \frac{{\left( {{{\vec{T}}_\mathrm A} - {{\vec{N}}_\mathrm{AB}}} \right) \cdot {{\vec{v}}_\mathrm A}}}{c} - \frac{{\left( {{n_\mathrm A} - 1} \right){{\vec{T}}_\mathrm A} \cdot {{\vec{v}}_\mathrm A}}}{c}}}\nonumber\\
 \approx \frac{{1 - \frac{{{{\vec{N}}_\mathrm{AB}} \cdot {{\vec{v}}_\mathrm B}}}{c} + \frac{{{v_\mathrm{Bx}}{\delta _\mathrm B}\sin {\gamma _\mathrm B} - {v_\mathrm{By}}{\delta _\mathrm B}\cos {\gamma _\mathrm B}}}{c} - \frac{{\left( {{M_1} + {M_2}} \right){{\vec{N}}_\mathrm{AB}} \cdot {{\vec{v}}_\mathrm B}}}{c}}}{{1 - \frac{{{{\vec{N}}_\mathrm{AB}} \cdot {{\vec{v}}_\mathrm A}}}{c} - \frac{{{v_\mathrm{Ax}}{\delta _\mathrm A}\sin {\gamma _\mathrm B} - {v_\mathrm{Ay}}{\delta _\mathrm A}\cos {\gamma _\mathrm B}}}{c} + \frac{{40.3{n_\mathrm e}{{\vec{N}}_\mathrm{AB}} \cdot {{\vec{v}}_\mathrm A}}}{{c{f^2}}}}}
\end{eqnarray}
Here, $v_\mathrm{Ax}$, $v_\mathrm{Ay}$, $v_\mathrm{Bx}$ and $v_\mathrm{By}$ are derived from the components of velocities in LLCS, which are determined by (\ref{vllcsa}) and (\ref{en}).

\section{}
\label{sec:b}
Considering ACES links $f_1$ and $f_2$, we mark the positions of the ground station at $t_1$ as $B^{\prime}$, space station at $t_2$ as $A^{\prime}$, space station at $t_3$ as $A$, and ground station at $t_5$ as $B$ (as figure \ref{fig2} shows).

To express the Doppler frequency shift to the order of $c^{-1}$, it is easy to obtain the shift of two links as
\begin{eqnarray}\label{dfdop12}
&\delta {f_\mathrm{dop1}} = {{\left( {1 - \frac{{{{\vec{N}}_\mathrm{B'A'}} \cdot {{\vec{v}}_\mathrm{A'}}}}{c}} \right)} \mathord{\left/
 {\vphantom {{\left( {1 - \frac{{{{\vec{N}}_\mathrm{B'A'}} \cdot {{\vec{v}}_\mathrm{A'}}}}{c}} \right)} {\left( {1 - \frac{{{{\vec{N}}_\mathrm{B'A'}} \cdot {{\vec{v}}_\mathrm{B'}}}}{c}} \right)}}} \right.
 \kern-\nulldelimiterspace} {\left( {1 - \frac{{{{\vec{N}}_\mathrm{B'A'}} \cdot {{\vec{v}}_\mathrm{B'}}}}{c}} \right)}}\nonumber\\
&\delta {f_\mathrm{dop2}} = {{\left( {1 - \frac{{{{\vec{N}}_\mathrm{AB}} \cdot {{\vec{v}}_\mathrm B}}}{c}} \right)} \mathord{\left/
 {\vphantom {{\left( {1 - \frac{{{{\vec{N}}_\mathrm{AB}} \cdot {{\vec{v}}_\mathrm B}}}{c}} \right)} {\left( {1 - \frac{{{{\vec{N}}_\mathrm{AB}} \cdot {{\vec{v}}_\mathrm A}}}{c}} \right)}}} \right.
 \kern-\nulldelimiterspace} {\left( {1 - \frac{{{{\vec{N}}_\mathrm{AB}} \cdot {{\vec{v}}_\mathrm A}}}{c}} \right)}}
\end{eqnarray}

Considering that the time interval of this process is very short (known for the height of ISS, which is approximately $T_{12}=T_{35}=1\ \mathrm{ms}$; according to \cite{Duchayne2009}, we need $T_{23}<1\ \mathrm{\mu s}$, two Doppler frequency shifts are numerically close.

To simplify the overall frequency shift, we must express all terms at time $t_1$ and $t_2$ by the terms at time $t_3$ and $t_5$. For this study, we deduced the algorithm adopted by the appendix of \cite{Blanchet2001}. Using this algorithm, we have
\begin{equation}\label{nbaprime}
{{\vec{N}}_\mathrm{B'A'}} = \left( { - {{\vec{N}}_\mathrm{AB}} + \frac{{{{\vec{r}}_\mathrm{A'}} - {{\vec{r}}_\mathrm A}}}{{{R_\mathrm{AB}}}} - \frac{{{{\vec{r}}_\mathrm{B'}} - {{\vec{r}}_\mathrm B}}}{{{R_\mathrm{AB}}}}} \right)\frac{{{R_\mathrm{AB}}}}{{{R_\mathrm{A'B'}}}}
\end{equation}
\begin{equation}\label{raprime}
{{\vec{r}}_\mathrm{A'}} - {{\vec{r}}_\mathrm A} =  - {{\vec{v}}_\mathrm A}{T_{23}} + \frac{1}{2}{{\vec{a}}_\mathrm A}T_{23}^2
\end{equation}
\begin{equation}\label{rbprime}
{{\vec{r}}_\mathrm{B'}} - {{\vec{r}}_\mathrm B} =  - {{\vec{v}}_\mathrm B}{T_{15}} + \frac{1}{2}{{\vec{a}}_\mathrm B}T_{15}^2
\end{equation}
\begin{equation}\label{t15}
{T_{15}} = \frac{{2{R_\mathrm{AB}}}}{c} - \frac{{2{{\vec{R}}_\mathrm{AB}} \cdot {{\vec{v}}_\mathrm B}}}{{{c^2}}} + {T_{23}}
\end{equation}

According to the actual value, we roughly calculated the speed value of ISS, and $v/c$ is approximately $10^{-5}$. Based on this magnitude, the magnitude of $T_{23}$ is approximately nanosecond; thus, they can be considered a $c^{-2}$ term. However, $T_{15}$ is considered a $c^{-1}$ term according to equation (\ref{t15}). By equations (\ref{raprime}) and (\ref{rbprime}), neglecting $O(c^{-3})$, we can obtain vector $R_\mathrm{A^{\prime}B^{\prime}}$
\begin{equation}\label{rabprime}
{{\vec{R}}_\mathrm{A'B'}} = {{\vec{R}}_\mathrm{AB}} + {{\vec{v}}_\mathrm A}{T_{23}} - {{\vec{v}}_\mathrm B}{T_{15}} + \frac{1}{2}{{\vec{a}}_\mathrm B}T_{15}^2
\end{equation}
After squaring the vector and expanding the root of this result into secondary series, we obtain:
\begin{eqnarray}\label{rabprime2}
{R_\mathrm{A'B'}} &= {R_\mathrm{AB}} - \frac{{\left( {{{\vec{R}}_\mathrm{AB}} \cdot {{\vec{v}}_\mathrm B}} \right){T_{15}}}}{{{R_\mathrm{AB}}}} + \frac{{\left( {{{\vec{R}}_\mathrm{AB}} \cdot {{\vec{a}}_\mathrm B}} \right)T_{15}^2}}{{2{R_\mathrm{AB}}}} + \frac{{v_\mathrm B^2T_{15}^2}}{{2{R_\mathrm{AB}}}} \nonumber\\
&+ \frac{{\left( {{{\vec{R}}_\mathrm{AB}} \cdot {{\vec{v}}_\mathrm A}} \right){T_{23}}}}{{{R_\mathrm{AB}}}} - \frac{{{{\left( {{{\vec{R}}_\mathrm{AB}} \cdot {{\vec{v}}_\mathrm B}} \right)}^2}T_{15}^2}}{{2R_\mathrm{AB}^3}}
\end{eqnarray}
Substitute $T_{15}$ with (\ref{t15}), we have
\begin{eqnarray}\label{rabprime3}
{R_\mathrm{A'B'}} &= {R_\mathrm{AB}} - \frac{{2{{\vec{R}}_\mathrm{AB}} \cdot {{\vec{v}}_\mathrm B}}}{c} + \frac{{2\left( {{{\vec{R}}_\mathrm{AB}} \cdot {{\vec{a}}_\mathrm B}} \right){R_\mathrm{AB}}}}{{{c^2}}} + \frac{{2v_\mathrm B^2{R_\mathrm{AB}}}}{{{c^2}}}\nonumber\\
 &- \frac{{{{\vec{R}}_\mathrm{AB}} \cdot \left( {{{\vec{v}}_\mathrm B} - {{\vec{v}}_\mathrm A}} \right)}}{{{R_\mathrm{AB}}}}{T_{23}}
\end{eqnarray}
and
\begin{eqnarray}\label{rab_rabprime}
\frac{{{R_\mathrm{AB}}}}{{{R_\mathrm{A'B'}}}} &= 1 - \frac{{{R_\mathrm{A'B'}} - {R_\mathrm{AB}}}}{{{R_\mathrm{AB}}}} + {\left( {\frac{{{R_\mathrm{A'B'}} - {R_\mathrm{AB}}}}{{{R_\mathrm{AB}}}}} \right)^2}\nonumber\\
&= 1 + \frac{{2{{\vec{N}}_\mathrm{AB}} \cdot {{\vec{v}}_\mathrm B}}}{c} - \frac{{2\left( {{{\vec{N}}_\mathrm{AB}} \cdot {{\vec{a}}_\mathrm B}} \right){R_\mathrm{AB}}}}{{{c^2}}} - \frac{{2v_\mathrm B^2}}{{{c^2}}} + \frac{{{{\vec{N}}_\mathrm{AB}} \cdot \left( {{{\vec{v}}_\mathrm B} - {{\vec{v}}_\mathrm A}} \right)}}{{{R_\mathrm{AB}}}}{T_{23}}\nonumber \\
&+ \frac{{4{{\left( {{{\vec{N}}_\mathrm{AB}} \cdot {{\vec{v}}_\mathrm B}} \right)}^2}}}{{{c^2}}}
\end{eqnarray}
From (\ref{nbaprime})-(\ref{rbprime}) and (\ref{rab_rabprime}), we obtain
\begin{eqnarray}\label{nabprime2}
{{\vec{N}}_\mathrm{B'A'}} &=  - {{\vec{N}}_\mathrm{AB}}\left[ 1 + \frac{{2{{\vec{N}}_\mathrm{AB}} \cdot {{\vec{v}}_\mathrm B}}}{c} - \frac{{2\left( {{{\vec{N}}_\mathrm{AB}} \cdot {{\vec{a}}_\mathrm B}} \right){R_\mathrm{AB}}}}{{{c^2}}} - \frac{{2v_\mathrm B^2}}{{{c^2}}} +\right.\nonumber\\ 
&\left.\frac{{{{\vec{N}}_\mathrm{AB}} \cdot \left( {{{\vec{v}}_\mathrm B} - {{\vec{v}}_\mathrm A}} \right)}}{{{R_\mathrm{AB}}}}{T_{23}} + \frac{{4{{\left( {{{\vec{N}}_\mathrm{AB}} \cdot {{\vec{v}}_\mathrm B}} \right)}^2}}}{{{c^2}}} \right]\nonumber\\
&+ \frac{2}{c}{{\vec{v}}_\mathrm B}\left( {1 + \frac{{{{\vec{N}}_\mathrm{AB}} \cdot {{\vec{v}}_\mathrm B}}}{c}} \right) + \frac{{{T_{23}}}}{{{R_\mathrm{AB}}}}\left( {{{\vec{v}}_\mathrm B} - {{\vec{v}}_\mathrm A}} \right) - \frac{2}{{{c^2}}}{R_\mathrm{AB}}{{\vec{a}}_\mathrm B}
\end{eqnarray}
With the velocity of the ground station and space station, in terms of velocity, acceleration and derivative of acceleration \cite{Blanchet2001}, we have
\begin{equation}\label{vaprime}
{{\vec{v}}_\mathrm{A'}} = {{\vec{v}}_\mathrm A} - {{\vec{a}}_\mathrm A}{T_{23}}
\end{equation}
\begin{equation}\label{vbprime}
{{\vec{v}}_\mathrm{B'}} = {{\vec{v}}_\mathrm B} - \frac{2}{c}{R_\mathrm{AB}}{{\vec{a}}_\mathrm B} + \frac{2}{{{c^2}}}\left[ {\left( {{R_\mathrm{AB}} \cdot {v_\mathrm B}} \right){{\vec{a}}_\mathrm B} + {R_\mathrm{AB}}{{\vec{b}}_\mathrm B}} \right] - {{\vec{a}}_\mathrm B}{T_{23}}
\end{equation}
For the approximation of ACES: $\left| {{{\vec{N}}_\mathrm{AB}} \cdot {{\vec{v}}_\mathrm A}/c} \right| \le 2.6 \times {10^{ - 5}}$, $\left| {{{\vec{N}}_\mathrm{AB}} \cdot {{\vec{v}}_\mathrm B}/c} \right| \le 1.6 \times {10^{ - 6}}$, $\left| {{{\vec{R}}_\mathrm{AB}} \cdot {{\vec{a}}_\mathrm A}/{c^2}} \right| \le 1.7 \times {10^{ - 10}}$, and $\left| {{{\vec{R}}_\mathrm{AB}} \cdot {{\vec{a}}_\mathrm B}/{c^2}} \right| \le 7 \times {10^{ - 13}}$ \cite{Blanchet2001}; to the order of $10^{-16}$, we have
\begin{eqnarray}\label{nva}
\fl 1 - \frac{{{{\vec{N}}_\mathrm{B'A'}} \cdot {{\vec{v}}_\mathrm{A'}}}}{c} &= 1 + \frac{{{{\vec{N}}_\mathrm{AB}} \cdot {{\vec{v}}_\mathrm A}}}{c} + \frac{{2\left( {{{\vec{N}}_\mathrm{AB}} \cdot {{\vec{v}}_\mathrm A}} \right)\left( {{{\vec{N}}_\mathrm{AB}} \cdot {{\vec{v}}_\mathrm B}} \right)}}{{{c^2}}} - \frac{{2{{\vec{v}}_\mathrm A} \cdot {{\vec{v}}_\mathrm B}}}{{{c^2}}}\nonumber\\
&+\left[ {\frac{{{{\vec{N}}_\mathrm{AB}} \cdot \left( {{{\vec{v}}_\mathrm B} - {{\vec{v}}_\mathrm A}} \right)\left( {{{\vec{N}}_\mathrm{AB}} \cdot {{\vec{v}}_\mathrm A}} \right)}}{{c{R_\mathrm{AB}}}} - \frac{{\left( {{{\vec{v}}_\mathrm B} - {{\vec{v}}_\mathrm A}} \right) \cdot {{\vec{v}}_\mathrm A}}}{{c{R_\mathrm{AB}}}} - \frac{{{{\vec{N}}_\mathrm{AB}} \cdot {{\vec{a}}_\mathrm A}}}{c}} \right]{T_{23}}
\end{eqnarray}
\begin{eqnarray}\label{nvb}
\fl 1 - \frac{{{{\vec{N}}_\mathrm{B'A'}} \cdot {{\vec{v}}_\mathrm{B'}}}}{c} &= 1 + \frac{{{{\vec{N}}_\mathrm{AB}} \cdot {{\vec{v}}_\mathrm B}}}{c} + \frac{{2{{\left( {{{\vec{N}}_\mathrm{AB}} \cdot {{\vec{v}}_\mathrm B}} \right)}^2}}}{{{c^2}}} - \frac{{2v_\mathrm B^2}}{{{c^2}}} - \frac{{2{R_\mathrm{AB}}{{\vec{N}}_\mathrm{AB}} \cdot {{\vec{a}}_\mathrm B}}}{{{c^2}}}\nonumber\\
&+\left[ {\frac{{{{\vec{N}}_\mathrm{AB}} \cdot \left( {{{\vec{v}}_\mathrm B} - {{\vec{v}}_\mathrm A}} \right)\left( {{{\vec{N}}_\mathrm{AB}} \cdot {{\vec{v}}_\mathrm B}} \right)}}{{c{R_\mathrm{AB}}}} - \frac{{\left( {{{\vec{v}}_\mathrm B} - {{\vec{v}}_\mathrm A}} \right) \cdot {{\vec{v}}_\mathrm B}}}{{c{R_\mathrm{AB}}}}} \right]{T_{23}}
\end{eqnarray}

\section*{References}
\bibliographystyle{iopart-num}
\bibliography{ref}

\end{document}